\documentclass[twocolumn]{aastex631}   

\usepackage{newtxtext,newtxmath}
\usepackage[T1]{fontenc}

\usepackage{nicefrac}
\usepackage{amssymb}
\usepackage{latexsym}
\usepackage{amsmath, mathrsfs, bm}
\usepackage{eqnarray}
\usepackage{url}
\usepackage{cases}
\usepackage{epsfig}
\usepackage{color}
\usepackage{graphicx}
\usepackage{grffile}
\usepackage{float}
\usepackage{soul}
\usepackage{enumerate}
\usepackage{multirow}
\usepackage{threeparttable}
\usepackage{xspace}

\usepackage{bm}
\usepackage{times}

\usepackage{etoolbox}
\usepackage{multirow}

\begin{document}

\title{Dark Energy in the DESI Era: A Brief Review of Evidence, Beyond-$\Lambda$CDM Interpretations, and Tensions}

\author[0009-0004-6982-4021]{Tian-Nuo Li}
\affiliation{Liaoning Key Laboratory of Cosmology and Astrophysics, College of Sciences, Northeastern University, Shenyang 110819, China}

\author[0009-0005-6921-3201]{Guo-Hong Du}
\affiliation{Liaoning Key Laboratory of Cosmology and Astrophysics, College of Sciences, Northeastern University, Shenyang 110819, China}

\author[0000-0003-0232-8344]{Hao Wang}
\affiliation{School of Fundamental Physics and Mathematical Sciences, Hangzhou Institute for Advanced Study, UCAS, Hangzhou 310024, China}
\affiliation{School of Physical Sciences, University of Chinese Academy of Sciences, Beijing 100049, China}

\author[0009-0004-4056-5775]{Yun-He Li}
\affiliation{Liaoning Key Laboratory of Cosmology and Astrophysics, College of Sciences, Northeastern University, Shenyang 110819, China}

\author[0000-0002-3512-2804]{Jing-Fei Zhang}
\affiliation{Liaoning Key Laboratory of Cosmology and Astrophysics, College of Sciences, Northeastern University, Shenyang 110819, China}

\author[0000-0002-6029-1933]{Xin Zhang}
\affiliation{Liaoning Key Laboratory of Cosmology and Astrophysics, College of Sciences, Northeastern University, Shenyang 110819, China}
\affiliation{MOE Key Laboratory of Data Analytics and Optimization for Smart Industry, Northeastern University, Shenyang 110819, China}
\affiliation{National Frontiers Science Center for Industrial Intelligence and Systems Optimization, Northeastern University, Shenyang 110819, China}


\correspondingauthor{Xin Zhang}
\email{zhangxin@mail.neu.edu.cn}

\begin{abstract}

Recent baryon acoustic oscillation measurements from Dark Energy Spectroscopic Instrument (DESI) provide important new clues for reassessing whether the standard $\Lambda$CDM model offers a sufficient description of the late-time expansion history of the universe. When combined with cosmic microwave background and type Ia supernova data, these measurements show an apparent departure from the $\Lambda$CDM model, commonly described as dynamical dark energy (DDE) with equation of state crossing the phantom divide (i.e., quintom behavior). This review examines the current status of the DESI-motivated indications for DDE and their possible implications for physics beyond $\Lambda$CDM. We discuss how the strength of the preference for DDE depends on the adopted parameterization and dataset combination, and how residual systematics or internal tensions among datasets may affect its interpretation. At the background level, several mechanisms beyond $\Lambda$CDM can produce similar expansion histories. We therefore further discuss how the same effective departure from $w=-1$ may arise from physically distinct scenarios, including interacting dark energy, non-minimally coupled gravity, and non-standard dark matter. Meanwhile, these different new-physics interpretations may have different implications for current cosmological tensions, especially those involving $H_0$, $S_8$, and $\sum m_\nu$. In conclusion, the question posed by DESI is not merely whether dark energy evolves with time, but rather how, within the framework of precision cosmology, to disentangle new physics scenarios from systematic errors.

\end{abstract}


\section{Introduction}

The late-time accelerated expansion of the universe remains one of the most profound open questions in modern cosmology and fundamental physics. Since its discovery through observations of distant type Ia supernovae (SNe)~\citep{SupernovaSearchTeam:1998fmf,SupernovaCosmologyProject:1998vns}, cosmic acceleration has been independently supported by a wide range of cosmological probes, including cosmic microwave background (CMB) \citep{WMAP:2003elm,WMAP:2003ivt,Planck:2018vyg} and baryon acoustic oscillation (BAO) \citep{SDSS:2005xqv,BOSS:2016wmc,eBOSS:2020yzd}. Within the framework of general relativity, such an accelerated expansion requires a component with sufficiently negative pressure, usually described by an equation of state (EoS) satisfying $w<-1/3$. This unknown component, commonly referred to as dark energy (DE), dominates the current energy budget of the universe. However, its physical origin remains unclear. It may correspond to vacuum energy, a new dynamical field, or an indication that gravity itself is modified on cosmological scales.

The simplest and most successful phenomenological description of DE is provided by the cosmological constant $\Lambda$, which, together with cold dark matter (CDM), forms the standard $\Lambda$CDM model. Despite its minimal number of parameters, $\Lambda$CDM has achieved remarkable success in explaining a broad range of observations, from the CMB angular power spectra to BAO measurements, SNe, and the growth of cosmic structure~\citep{Weinberg:2013agg,Huterer:2017buf}. Nevertheless, the physical interpretation of the cosmological constant is far from satisfactory. The enormous discrepancy between the observed value of the vacuum energy density and theoretical expectations from quantum field theory gives rise to the well-known fine-tuning problem~\citep{Sahni:1999gb}, while the comparable energy densities of dark matter and DE today motivate the cosmic coincidence problem~\citep{Zlatev:1998tr,Dalal:2001dt}. In addition, the persistent $H_0$ tension between early- and late-universe measurements \citep{Li:2013dha,Zhao:2017urm,Guo:2018ans,Verde:2019ivm,Planck:2018vyg,Vagnozzi:2019ezj,DiValentino:2020zio,DiValentino:2021izs,Shah:2021onj,Vagnozzi:2021gjh,Gao:2021xnk,Riess:2021jrx,Perivolaropoulos:2021jda,Schoneberg:2021qvd,Abdalla:2022yfr,DiValentino:2022fjm,Kamionkowski:2022pkx,Giare:2023xoc,Hu:2023jqc,Vagnozzi:2023nrq,Cai:2026swf}, together with the $S_8$ tension between CMB-inferred clustering amplitudes and low-redshift large-scale-structure observations~\citep{Planck:2018vyg,KiDS:2020suj,DiValentino:2020vvd,Abdalla:2022yfr,Wright:2025xka,Pantos:2026koc}, has raised the possibility that $\Lambda$CDM may be an effective description rather than the final cosmological model.

At the theoretical level, a broad and well-studied possibility is that cosmic acceleration is driven by additional dynamical degrees of freedom, most commonly represented by scalar fields~\citep{Ratra:1987rm,Copeland:2006wr}. In contrast to the cosmological constant, whose EoS is fixed at $w=-1$, a scalar field generally evolves with cosmic time, so that its kinetic and potential contributions lead to an effective DE density and pressure that need not remain constant. The simplest realization is canonical quintessence~\citep{Ratra:1987rm,Caldwell:1997ii,Zlatev:1998tr,Zhang:2006av,Zhang:2008mb,Tsujikawa:2013fta}, for which the EoS typically satisfies $w>-1$. More general constructions, such as phantom fields with $w<-1$~\citep{Caldwell:1999ew}, quintom models with $w$ crossing $-1$~\citep{Feng:2004ad,Guo:2004fq,Zhang:2005kj,Zhang:2005yz,Zhang:2005hs,Zhang:2006qu,Ma:2007av,Cai:2009zp,Li:2012via,Cai:2025mas}, and k-essence~\citep{Armendariz-Picon:2000nqq}, can give rise to richer dynamical behavior.

Although these models differ substantially in their underlying Lagrangians, field dynamics, and stability properties, their impact on the homogeneous expansion history can often be summarized by a time-dependent EoS of DE at an effective level. This motivates the use of phenomenological parametrizations of $w(a)$ in observational analyses. Such parametrizations should not be regarded as unique microscopic models of DE, but rather as theory-agnostic descriptions of possible departures from the cosmological-constant limit in the background expansion. They therefore provide a useful first step for assessing whether current data prefer a dynamical dark energy (DDE) component before committing to a specific theoretical realization. A widely used example is the Chevallier-Polarski-Linder (CPL) parameterization \citep{Chevallier:2000qy,Linder:2002et},
\begin{equation}
    w(a)=w_0+w_a(1-a),\label{CPL}
\end{equation}
where $w_0$ denotes the present-day value of the EoS of DE and $w_a$ characterizes its leading-order time variation. Recent BAO measurements from the Dark Energy Spectroscopic Instrument (DESI) have brought renewed attention to this class of models. When DESI BAO data are combined with CMB and SN data, current analyses report a preference for a present-day quintessence-like EoS, $w_0>-1$, together with a negative evolution parameter, $w_a<0$. In the CPL framework, this corresponds to an evolution in which DE was more phantom-like in the past and becomes quintessence-like at the present epoch, implying a possible crossing of the phantom divide ($w=-1$). Depending on the SN compilation and dataset combination employed, this preference reaches a statistical significance in the range $2.8-3.8\sigma$~\citep{DESI:2024mwx,DESI:2025zgx,DES:2025sig}.

However, the interpretation of this apparent preference for DDE requires caution. Since the evolution of DE is not measured directly, but inferred through an assumed phenomenological form of $w(a)$, the result may depend on the adopted parameterization. Different choices of $w(a)$ can lead to different reconstructions of the late-time expansion history and may alter the statistical significance of the departure from $\Lambda$CDM \citep{Jassal:2005qc,Barboza:2008rh,Ma:2011nc,Dimakis:2016mip,Pan:2019brc}. Therefore, testing the parameterization dependence of the DESI-motivated signal is a necessary step in assessing whether it reflects a genuine feature of the data or an artifact of a specific ansatz such as CPL~\citep{Giare:2024gpk,DESI:2025fii,Ormondroyd:2025iaf,Li:2025vuh}. Equally important is the dependence on the observational datasets themselves. The strength of the preference depends sensitively on the choice of data combinations~\citep{Giare:2025pzu}. In particular, DESY5 SN data tend to enhance the evidence for DDE, whereas combinations involving PantheonPlus, Sloan Digital Sky Survey (SDSS) BAO, or non-Planck CMB measurements can weaken it~\citep{Ghosh:2024kyd,Giare:2024oil,Xu:2026sbw}. Possible systematics in DESI BAO measurements at specific redshifts, calibration differences among SN samples, and the role of large-scale Planck temperature and polarization data may all influence the inferred deviation from $\Lambda$CDM~\citep{Efstathiou:2024xcq}. In this sense, the current evidence is best understood as an intriguing but not yet conclusive indication of DDE, whose robustness must be evaluated against both parameterization dependence and data-related systematics. 

Beyond the question of robustness, the DESI-motivated departure from  $\Lambda$CDM also raises a deeper question of physical interpretation. Although this departure is commonly described phenomenologically as an evolution of the EoS of DE, such a description does not uniquely identify conventional DDE as the underlying mechanism. Late-time probes such as BAO and SN are primarily sensitive to the redshift evolution of the cosmic expansion rate, or equivalently to the energy-density evolution that sources it \citep{Weinberg:2013agg}. Therefore, similar departures from $\Lambda$CDM may arise if the dark sectors are not separately conserved \citep{Amendola:1999er,Farrar:2003uw,Das:2005yj}. For example, in interacting dark energy (IDE) models, energy--momentum exchange between dark matter and DE changes the evolution of the dark-sector densities, so that an apparent evolving EoS may be inferred even when the intrinsic DE component is not a standard dynamical fluid \citep{Zhang:2005rg,Avelino:2012tc,Li:2013bya,Li:2014eha,Wang:2016lxa,Giare:2024smz,Li:2024qso,Li:2025owk,Li:2026xaz}.

Another possibility is that the apparent deviation is not produced by an independently conserved DE fluid, but by a different way in which cosmic acceleration is encoded in the field equations \citep{Carroll:2003wy,Clifton:2011jh}. In non-minimally coupled scalar-field scenarios, the relation among matter, spacetime geometry, and the scalar degree of freedom differs from that in minimally coupled DE within general relativity \citep{Gubitosi:2012hu,Koyama:2015vza,Cai:2015emx,Ishak:2018his}. When these theories are rewritten in an effective general-relativistic form, the additional terms can be absorbed into an effective DE density and pressure, leading to an apparent evolution of $w(a)$ \citep{Chudaykin:2024gol,Yang:2024kdo,Ye:2024ywg,Taule:2024bot,Ishak:2024jhs,Wolf:2025jed,Pan:2025psn}.

Moreover, a departure from $\Lambda$CDM does not necessarily have to be attributed to DE. It may also reflect limitations of the standard CDM assumption. Although CDM is highly successful on cosmological and large scales, it faces long-standing small-scale challenges \citep{Bullock:2017xww,Tulin:2017ara}, which have motivated a variety of non-standard dark matter scenarios. If dark matter is not exactly cold and pressureless, as characterized by a non-zero EoS parameter ($w_{\rm dm}$), its density evolution and perturbation growth would differ from those of standard CDM \citep{Muller:2004yb,Kopp:2018zxp,Ilic:2020onu}. Such changes can affect the inferred expansion history and structure growth, thereby altering the interpretation of the apparent departure from $\Lambda$CDM \citep{Kumar:2025etf,Yang:2025ume,Abedin:2025dis,Yang:2025boq,Chen:2025wwn,Yao:2025kuz,Li:2025dwz,Li:2025eqh,Braglia:2025gdo}.

It is also crucial to address the existing tensions in cosmological parameter measurements. First, while the DESI results imply DDE, they still fail to alleviate the well-known $H_0$ tension~\citep{DESI:2025zgx,Pang:2025lvh,Wang:2024dka}. This is because the EoS of DE that evolves into a quintessence-like regime at the present epoch typically yields a lower inferred $H_0$, thereby further widening the discrepancy with local distance-ladder measurements. Furthermore, the long-standing $S_8$ tension remains a critical issue that these extended models must confront~\citep{Wright:2025xka,Pantos:2026koc}. Finally, the inclusion of DESI data has sparked a striking new tension regarding the measurement of the total neutrino mass ($\sum m_\nu$). The remarkably tight upper bounds on $\sum m_\nu$ derived from recent cosmological constraints conflict with the established lower limits measured by terrestrial particle physics and neutrino oscillation experiments, posing a profound puzzle at the intersection of cosmology and fundamental physics~\citep{Zhang:2015uhk,Zhao:2016ecj,Zhang:2017rbg,Vagnozzi:2018jhn,Esteban:2020cvm,deSalas:2020pgw,KATRIN:2024cdt,DESI:2025zgx,DESI:2025ffm,Elbers:2025vlz,Du:2025xes,Giare:2025ath,Namikawa:2025doa,Sailer:2025lxj}.

With these considerations in mind, this review aims to clarify the status and interpretation of the recent DESI-motivated hints for deviations from $\Lambda$CDM. We discuss whether the apparent preference for DDE remains robust under variations in parameterization choices, dataset selection, and possible residual systematic effects. We also stress that an effective departure from $w=-1$ at the background level does not uniquely identify the underlying physics. The same signal may be associated with DDE, but it may also be reinterpreted within IDE scenarios, non-minimally coupled gravities, or non-standard dark matter frameworks. We further examine how these different interpretations impact the main cosmological tensions, especially those related to $H_0$, $S_8$, and $\sum m_\nu$. In this way, the DESI results provide not only a possible indication of DDE, but also a timely opportunity to reassess the assumptions underlying the standard cosmological model. Determining whether these hints point to new physics or to the combined effect of parameterization choices, dataset combinations, and systematic uncertainties is therefore a central task for precision cosmology.

\section{Evidence for Dynamical Dark Energy from DESI}\label{sec:data_setting}

Recent measurements from DESI have revived considerable interest in the possibility of DDE \citep{DESI:2024mwx,DESI:2025zgx}. When interpreted within phenomenological parametrizations of the EoS of DE, combinations of DESI BAO measurements with CMB and SN observations reveal intriguing hints of departures from the $\Lambda$CDM paradigm, making the robustness of this indication an important subject of ongoing investigation \citep{Li:2024qus,Liu:2025mub,Wu:2025wyk,Wang:2025vtw,Kibris:2026cqq,Anchordoqui:2026hys,Montefalcone:2026iga,vanderWesthuizen:2026ryk,Song:2025bio,Wolf:2025acj,Khandelwal:2026sgm}.

\subsection{Observational results from DESI}
\label{subsec:desi_evidence}

\begin{figure}[t]
\centering
\includegraphics[width=0.48\textwidth]{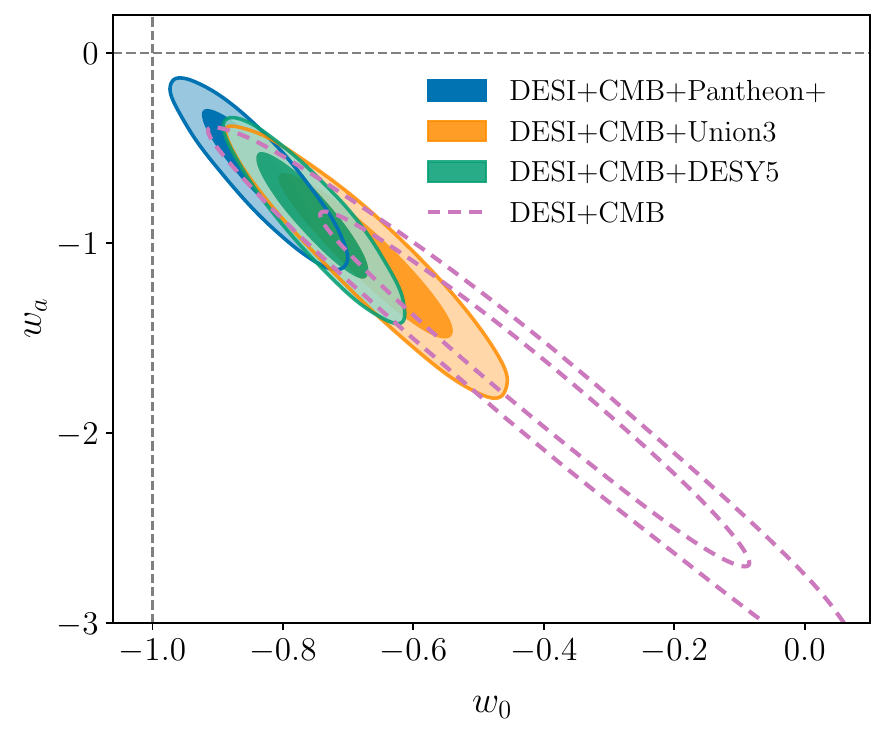}
\caption{
Two-dimensional marginalized contours for the CPL parameters $(w_0,w_a)$ from different combinations of DESI BAO, CMB, and SN datasets~\citep{DESI:2025zgx}. 
}
\label{fig1}
\end{figure}

In 2024 April, the DESI collaboration released its Data Release 1 (DR1) BAO data~\citep{DESI:2024mwx}. When DESI DR1 BAO measurements are combined with Planck CMB observations alone, the preference for DDE remains relatively weak. However, combined with specific SN compilations, the statistical significance of the DDE preference ranges approximately from $2.5\sigma$ to $3.9\sigma$ within the CPL framework, which suggests that low-redshift probes play a crucial role in driving the apparent preference for DDE. The DESI Data Release 2 (DR2) analyses further strengthened this trend up to 4.2$\sigma$~\citep{DESI:2025zgx}. Figure~\ref{fig1} illustrates the cosmological constraints in the $(w_0,w_a)$ plane obtained from different combinations of DESI BAO, CMB, and SN datasets. A notable feature is that once low-redshift probes are included, the preferred parameter regions systematically shift away from the $\Lambda$CDM point $(w_0,w_a)=(-1,0)$ toward the quadrant characterized by $w_0>-1, w_a<0$, indicating a nontrivial evolution of the DE sector. In particular, this preferred evolution typically corresponds to DE being more phantom-like at intermediate redshifts and approaching a quintessence-like state at the present epoch. The fact that different SN compilations qualitatively favor similar regions in parameter space suggests that the observed trend is unlikely to be entirely driven by a single dataset alone, although its precise statistical significance remains strongly dataset dependent. 

The evolution of the EoS of DE crossing the phantom divide is commonly referred to as the quintom scenario \citep{Feng:2004ad,Hu:2004kh,Guo:2004fq,Zhang:2005kj}, which cannot be achieved in conventional single-scalar-field models with canonical kinetic terms. Consequently, realizing a quintom evolution generally requires additional degrees of freedom, noncanonical scalar fields, higher-derivative operators, or effective descriptions beyond standard scalar-field DE. Beyond these field-theoretical approaches, quintom scenarios can also emerge in broader frameworks. It is worth emphasizing that the realization of a quintom scenario within the holographic DE framework was first pointed out by \citet{Zhang:2005yz}. In this framework, when the holographic parameter $c<1$, the EoS of the holographic DE can cross the phantom divide, making it behave as a quintom-type DE \citep{Zhang:2005yz,Zhang:2005hs}. Subsequently, \citet{Zhang:2006qu} introduced the notion of holographic quintom and further demonstrated that it can be effectively reconstructed by a generalized ghost condensate scalar-field model. Motivated by the DESI results, a large number of recent studies have revisited quintom-like DE evolution from both phenomenological and theoretical perspectives \citep{Yang:2024kdo,Goh:2025upc,Cai:2025mas,Gialamas:2025pwv,Ozulker:2025ehg,Yang:2025mws,Qiu:2025oop,Chen:2025ywv,Tsujikawa:2025wca,Abdalla:2026sis,Kang:2026vqj,Tsujikawa:2026xqm,Thanankullaphong:2026anl,Li:2026hwq,Shlivko:2026jxa,Gokcen:2026pkq,Wu:2025vfs}.

Nevertheless, despite the excitement generated by these results, it is important to emphasize that the current evidence for DDE remains far from conclusive. First, the current indication for DDE is fundamentally obtained within the CPL parameterization framework, while alternative parametrizations or nonparametric reconstructions of $w(a)$ may lead to quantitatively different conclusions. Second, the statistical significance of the deviation from $\Lambda$CDM depends sensitively on the adopted external datasets, particularly the treatment and calibration of SN samples. Recent studies have shown that different SN compilations can substantially alter the inferred significance of DDE, with some combinations significantly weakening the evidence for an evolving EoS. Finally, residual observational systematics and possible inconsistencies among cosmological datasets cannot yet be excluded.

\begin{figure*}[t]
\centering
\includegraphics[width=0.95\textwidth]{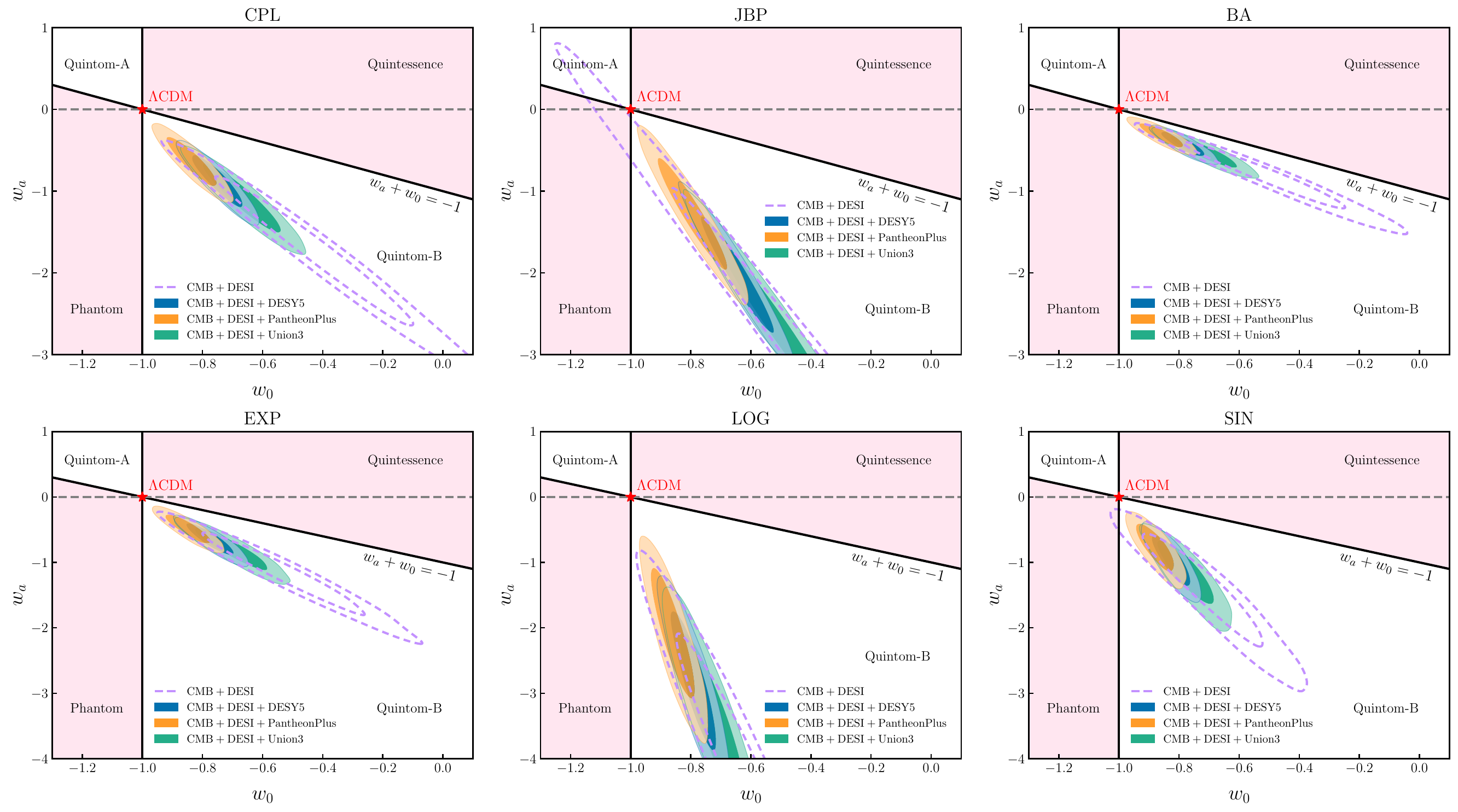}
\caption{Two-dimensional marginalized contours in the $(w_0, w_a)$ plane for six DDE parametrizations (CPL, JBP, BA, EXP, LOG, and SIN). The constraints are obtained using combinations of CMB, DESI DR2 BAO, and SN data. The red star denotes the $\Lambda$CDM model at $(w_0, w_a) = (-1, 0)$. The boundary lines, specifically the vertical solid line ($w_0 = -1$) and the oblique solid line ($w_a + w_0 = -1$, characterizing the EoS of DE in the early universe $a\to0$ to distinguish early phantom or early quintessence behaviors), delineate the regions corresponding to quintessence, phantom, and quintom evolution \citep{Li:2025vuh}.}
\label{fig2}
\end{figure*}

\subsection{Parameterization Dependence of the DDE Evidence}
\label{subsec:parametrization}

An important question concerning the DESI indication for DDE is whether the observed preference depends strongly on the assumed parameterization of the EoS of DE. Since the majority of the current analyses are performed within the CPL framework, it is natural to ask whether the apparent deviation from $\Lambda$CDM could simply be an artifact of this specific phenomenological assumption.

Motivated by this issue, several recent studies have systematically investigated the robustness of the DDE evidence under alternative parametrizations of $w(a)$ \citep{Giare:2024gpk,Malekjani:2024bgi,Zheng:2024qzi,DESI:2025fii,Chaudhary:2025vzy,Li:2025vuh}. Notably, an extended analysis by the DESI collaboration utilizing DESI DR2 BAO measurements demonstrated that extending $\Lambda$CDM to include alternative two-parameter models or employing non-parametric techniques consistently yields similar dynamical trends and phantom crossing behaviors~\citep{DESI:2025fii}. This indicates that the observed deviation is not merely a statistical artifact of the CPL framework assumption. For other analyses on parametric dependence of DDE; see, e.g., \citet{Wang:2024hwd,Yadav:2025vgo,Cheng:2025lod,Specogna:2025guo,RoyChoudhury:2025dhe,Kessler:2025kju,Cheng:2025yue,Alam:2025epg,Lee:2025pzo,Montefalcone:2026iga,Xu:2026sbw,Yadav:2026smi}.

Specifically, \citet{Li:2025vuh} analyzed the robustness of the evidence for DDE across six distinct parametrizations of $w(a)$: (i) CPL parameterization with $w(a) = w_0 + w_a(1-a)$, (ii) Jassal-Bagla-Padmanabhan (JBP) parameterization with $w(a) = w_0 + w_a a(1-a)$, (iii) Barboza-Alcaniz (BA) parameterization with $w(a) = w_0 + w_a \frac{1-a}{a^2 + (1-a)^2}$, (iv) Exponential (EXP) parameterization with $w(a) = (w_0 - w_a) + w_a \exp(1-a)$, (v) Logarithmic (LOG) parameterization with $w(a) = w_0 - w_a \left[ a \ln\left(1 + \frac{1}{a}\right) - \ln 2 \right]$, and (vi) Sinusoidal (SIN) parameterization with $w(a) = w_0 - w_a \left[ a \sin\left(\frac{1}{a}\right) - \sin 1 \right]$. Figure~\ref{fig2} summarizes the constraints in the $(w_0,w_a)$ plane obtained for these DDE parametrizations using combinations of DESI DR2 BAO, CMB, and SN observations. While the shapes and degeneracy directions of the confidence contours vary among different parametrizations, the preference for DDE does not disappear when the CPL assumption is relaxed. In several cases, alternative parametrizations even strengthen the deviation from $\Lambda$CDM. In particular, the BA parameterization often provides the largest improvement in the goodness-of-fit relative to $\Lambda$CDM and yields some of the strongest statistical preferences for DDE. At the same time, this study also reveals that the precise statistical significance of the DDE signal remains sensitive to the adopted dataset combination, especially the choice of SN compilation. In general, the DESY5 dataset tends to strengthen the preference for DDE, while PantheonPlus often leads to a comparatively weaker deviation from $\Lambda$CDM. In particular, the preferred parameter regions systematically lie in the quadrant $w_0>-1, w_a<0$ corresponding to a quintom-B type evolution in which the EoS of DE evolves from a phantom-like phase in the past toward a quintessence-like behavior at the present epoch, further motivating continued investigations into the physical origin of the observed deviation from $\Lambda$CDM.

\subsection{Impact of BAO and SN Data on the DDE Evidence}
\label{subsec:bao_sn_systematics}

The current preference for DDE should not be viewed as the result of a single observable, but rather as the outcome of a nontrivial interplay between BAO distance measurements, CMB temperature and polarization, and SN distance moduli. BAO data provide precise geometric information over a broad redshift range, while SN determine the relative luminosity-distance relation at low and intermediate redshifts. In the CPL framework, the DESI+CMB+SN analyses tend to shift the preferred parameter region away from the $\Lambda$CDM point $(w_0,w_a)=(-1,0)$ toward $w_0>-1$ and $w_a<0$. However, the detailed origin of this shift is more subtle than a simple global failure of $\Lambda$CDM.

\begin{figure}[t]
	\centering
	\includegraphics[width=0.52\textwidth]{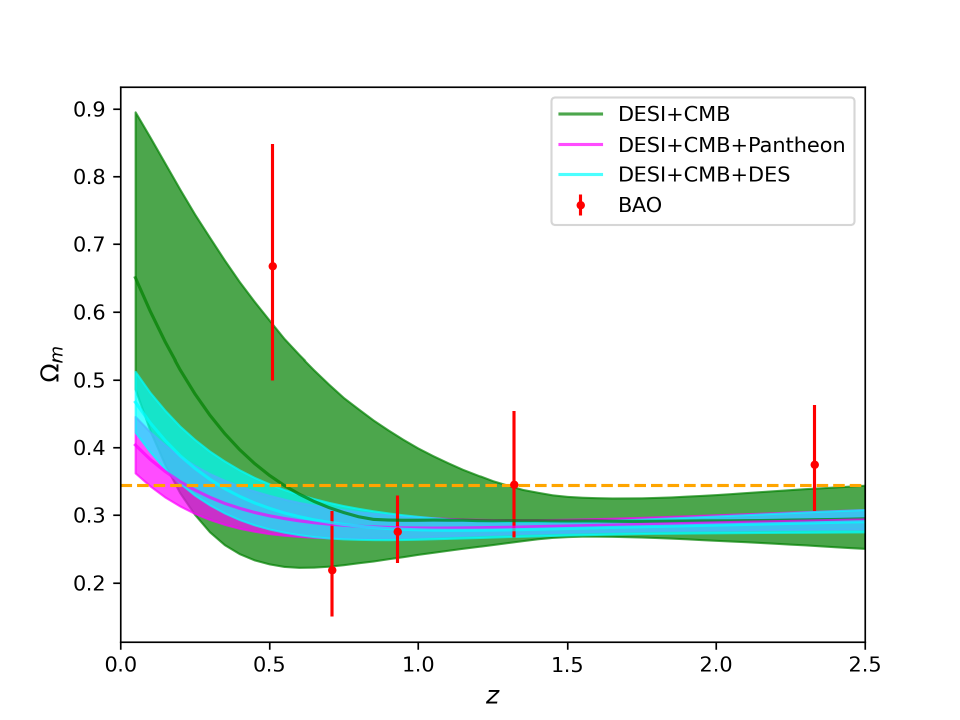}
	\caption{Inferred values of $\Omega_{\rm m}$ when the CPL cosmologies are mapped back to $\Lambda$CDM~\citep{Colgain:2024mtg}. The shaded intervals correspond to the $1\sigma$ confidence regions for the DESI+CMB (green), DESI+CMB+Pantheon (magenta), and DESI+CMB+DES (cyan) data. These are plotted alongside the constraints inferred from DESI anisotropic BAO measurements (red points).}
	\label{fig3}
\end{figure}

DESI BAO measurements play a central role because they constrain the transverse comoving distance $D_\mathrm{M}(z)/r_\mathrm{d}$ and the Hubble distance $D_\mathrm{H}(z)/r_\mathrm{d}$, thereby anchoring the late-time expansion history to the sound horizon scale inferred from the early universe. The official DESI analyses report good internal consistency between DR1 and DR2, as well as among the different BAO redshift bins, and therefore do not indicate a decisive internal failure of the BAO measurements \citep{DESI:2025zgx}. Nevertheless, several independent studies have emphasized that the contribution of BAO data to the DDE evidence is not uniformly distributed across redshift. In particular, the DESI DR1 luminous red galaxy (LRG) bins around $z_{\rm eff}\simeq 0.51$ and $z_{\rm eff}\simeq 0.71$ have been identified as especially influential. One analysis shows that the LRG bin at $z_{\rm eff}\simeq 0.51$ implies an unexpectedly high matter density, $\Omega_{\rm m}\simeq 0.67$, when interpreted within flat $\Lambda$CDM, and that this point drives the preference for $w_0>-1$ in DESI-only constraints \citep{Colgain:2024mtg}, as illustrated in Figure~\ref{fig3}. At the same time, the $z_{\rm eff}\simeq 0.71$ LRG bin tends to pull the inferred $\Omega_{\rm m}$ to lower values and has been discussed as an important contributor to the departure from the Planck-$\Lambda$CDM expectation \citep{Naredo-Tuero:2024sgf,Colgain:2024xqj}. 

Thus, although the full DESI BAO dataset is internally consistent at the current precision, a small number of intermediate-redshift BAO measurements can have a disproportionately large impact on the inferred DE parameters. A complementary way to isolate the origin of this effect is to recast the anisotropic BAO observables in terms of the angle-averaged distance $D_\mathrm{V}$ and the Alcock-Paczynski parameter $F_{\rm AP}=D_\mathrm{M}/D_\mathrm{H}$. Since $D_\mathrm{V}$ is more directly related to the monopole component of the galaxy clustering signal, whereas $F_{\rm AP}$ is more sensitive to anisotropic information, this decomposition allows one to identify which part of the BAO measurement is mainly responsible for the shift in cosmological parameters. Such an analysis finds that removing the $D_\mathrm{V}$ data point of the LRG2 bin shifts the $(w_0,w_a)$ contour closer to the $\Lambda$CDM limit and can bring the result into consistency with $w=-1$ within the $2\sigma$ contour \citep{Wang:2024rjd}. This suggests that monopole-related information in the LRG bins, especially LRG2, plays an important role in driving the apparent preference for DDE. These observations indicate that the DE inference from DESI is sensitive to specific redshift bins and to the way in which BAO information is compressed and combined with external datasets. The BAO data provide the geometric anchor for the deviation from $\Lambda$CDM, but the physical interpretation of this deviation requires caution, particularly because the same data points also influence other cosmological inferences, such as the neutrino-mass bound and the consistency between DESI and Planck \citep{Naredo-Tuero:2024sgf,Dinda:2024kjf,Wang:2024rjd,Wang:2024pui,Chudaykin:2024gol,Liu:2024gfy,Vilardi:2024cwq,Sapone:2024ltl}.


SNe play an even more direct role in determining the statistical strength of the DDE preference. While BAO measurements constrain distances relative to the sound horizon, SN provide the low-redshift luminosity-distance relation that is most sensitive to the late-time evolution of $w(a)$. This is why the evidence for DDE becomes substantially stronger when SN data are added to DESI BAO and CMB constraints. Conversely, the significance of the result depends strongly on which SN compilation is used. The PantheonPlus data generally gives a weaker preference for DDE, while Union3 and especially DESY5 tend to produce a stronger deviation from $\Lambda$CDM \citep{Efstathiou:2024xcq,DESI:2025zgx}. In this sense, the current DESI preference for DDE is as much a statement about the consistency between BAO, CMB, and SN distance scales as it is about DE itself. 

The DESY5 sample is particularly important because it yields one of the strongest preferences for DDE in the DESI combinations. However, several works have argued that the DESY5-driven enhancement may be sensitive to SN systematics. One concern is that DESY5 combines the high-redshift DES SN sample with an external low-redshift SN sample. The latter is heterogeneous, comes from multiple surveys, and has different calibration and selection properties. Even a relative offset of order a few hundredths of a magnitude between low- and high-redshift SN can significantly affect the inferred shape of the Hubble diagram and mimic the effect of an evolving EoS \citep{Efstathiou:2024xcq,Huang:2025som}. Indeed, analyses focusing on the DESY5 low-redshift anchor have argued that correcting for possible low-redshift SN systematics can reduce the preference for DDE to below the level usually regarded as significant \citep{Huang:2025som}. This does not prove that the original DESY5 result is wrong, but it shows that the interpretation of the DDE preference depends sensitively on the treatment of SN calibration and sample homogeneity. The DES-Dovekie recalibration provides an important independent test of this issue. This reanalysis updates the DESY5 calibration using improved photometric cross-calibration, recent white-dwarf calibration data, a retrained SALT3 light-curve model, and corrections to the treatment of the host-galaxy color law. After this recalibration, the inferred preference for DDE is reduced relative to the original DESY5 result: the reported significance decreases from about $4.2\sigma$ for DESY5 to about $3.2\sigma$ for DES-Dovekie when combined with DESI DR2 and CMB data \citep{DES:2025sig}. The evidence is therefore not eliminated, but its statistical strength is weakened, and the updated data exhibit only a weak Bayesian preference for CPL over $\Lambda$CDM.

Taken together, the current evidence for DDE is best understood as a dataset-dependent indication rather than a settled discovery. DESI BAO supplies the geometric tension, while SN data largely determine its statistical amplification. Future DESI releases, improved BAO consistency tests, independent low-redshift distance probes, and more robust SN recalibrations will be essential for determining whether the apparent preference for $w_0>-1$ and $w_a<0$ reflects genuine new physics or residual inconsistencies among current cosmological datasets.

\subsection{Reconstruction techniques for Dark Energy}
\label{subsec:nonparametric_pade_emergent}
Beyond the traditional CPL parameterization, recent studies have explored alternative approaches to reconstruct the evolution history of DE in a model-independent or phenomenologically motivated way. In general, reconstructions can be categorized into parametric and non-parametric approaches. The former imposes an explicit form on the quantities being reconstructed based on a priori assumption of a specific model, while non-parametric reconstructions gain purely data-driven insights into the dynamics of cosmological functions. These approaches offer complementary perspectives on the redshift evolution of the EoS of DE, collectively highlighting possible deviations from the cosmological constant.

Building on the model-independent insight, a flexible and systematic way is reconstructing cosmological functions by expanding them in terms of a chosen set of basis functions.
These methods include Taylor expansion~\citep{Visser:2003vq}, Fourier series~\citep{Tamayo:2019gqj}, wavelet expansion~\citep{Hojjati:2009ab} and Padé approximation~\citep{Gruber:2013wua,Wei:2013jya,Aviles:2014rma,Capozziello:2023ccw,Fazzari:2025lzd}. Particularly, Padé cosmography provides a complementary, high-precision method to characterize the late-time expansion. By expanding observables such as $H(z)$ and $D_\mathrm{L}(z)$ around multiple pivot redshifts using generalized Padé $(2,1)$ expansions, this approach mitigates truncation errors and avoids extrapolation biases. Importantly, the deviations inferred from Padé-based reconstructions, including cosmographic parameters $q(z)$ and $j(z)$, reinforce the earlier nonparametric findings: the generalized $Om(z)$ diagnostic shows $\sim 4\sigma$ tension relative to the constant-$w$ expectation \citep{Fazzari:2025lzd}. Thus, Padé cosmography confirms, through an independent mathematical strategy, that the expansion history implied by DESI and SN data is consistent with a DDE component at $z \lesssim 1$. See~\citet{Ling:2025lmw,Adil:2026kfn} for other reconstructions of cosmological functions.

Nonparametric and shape-function reconstructions utilize DESI DR1 /DR2 BAO measurements combined with SN and CMB priors to extract the redshift dependence of DE without assuming a specific parametric form~\citep{DESI:2025wyn}. Notably, they reveal mild oscillatory deviations from $w=-1$ at $z \lesssim 1$, with the DR2 dataset exhibiting a stronger dynamical signal than DR1. Moreover, derived shape functions $S_0(z)$, $S_1(z)$, and $S_2(z)$ systematically depart from $\Lambda$CDM expectations, consistently across multiple SN datasets. These results suggest that DE may possess internal degrees of freedom beyond a simple vacuum energy. As a powerful and versatile tool in machine learning, the Gaussian process has also been widely used in cosmology to reconstruct cosmological quantities in a model-independent way~\citep{Holsclaw:2010nb,Holsclaw:2010sk,Holsclaw:2011wi,Hwang:2022hla}, and the EoS of DE $w(a)$ has been widely reconstructed in redshift bins without imposing an explicit functional form as described in~\citet{Zhao:2012aw,Zhao:2017cud,Jiang:2024xnu,DESI:2025wyn,Abedin:2025yru,Ruchika:2025mkx,Cheng:2025cmb,Wang:2025xvi,Wang:2026kbg}.

\section{Beyond-$\Lambda$CDM interpretations}\label{sec:models}

Beyond the DDE interpretation discussed above, the DESI results also motivate a broader examination of possible new physics beyond the standard $\Lambda$CDM framework. If the apparent departure from $\Lambda$CDM indicated by current DESI data persists in future observations, it may point to the need to reassess some basic assumptions of the $\Lambda$CDM model. The central question is therefore what physical mechanism could underlie such an apparent departure. From this perspective, the post-DESI model-building effort should not be viewed simply as a search for alternative parametrizations of the EoS of DE. Rather, it is part of a broader attempt to identify which sector of the $\Lambda$CDM model may be responsible for the observed late-time deviation. 

This review primarily focuses on the following three broad classes of explanations:
\begin{enumerate}
    \item Interacting DE, in which DE and dark matter are allowed to exchange energy--momentum, thereby modifying both the background expansion history and the growth of cosmic structures;
    \item Non-minimally coupled gravity, in which a scalar field is non-minimally coupled to the Ricci scalar, providing an effective DDE component that can safely cross the phantom divide without introducing classical or quantum instabilities;
    \item Non-standard dark matter, which relaxes the standard assumption that dark matter is exactly cold and pressureless by allowing a non-zero EoS parameter of dark matter, $w_{\rm dm}\neq 0$. This may provide an alternative explanation for the apparent deviation from $\Lambda$CDM.
\end{enumerate}

These three directions should be viewed as complementary attempts to identify which assumption of the $\Lambda$CDM model is most directly implicated by the DESI-motivated departure. The same background expansion history can often be reproduced by modifying different sectors of the theory, but the physical interpretation is different in each case: the departure may reflect an interaction within the dark sector, a breakdown or extension of the general-relativistic description on cosmological scales, or a non-standard property of dark matter. The role of the following discussion is therefore to clarify how each class shifts the physical origin of the DESI preference.

\subsection{Interacting dark energy}\label{subsec:ide}

The DESI-motivated preference for DDE naturally raises a broader question: should the apparent departure from $\Lambda$CDM be attributed to the intrinsic dynamics of DE, or could it instead reflect a non-standard evolution of the dark sector as a whole? This distinction is essential because BAO measurements primarily constrain the background expansion history through distance-redshift relations and the Hubble expansion rate. They do not directly measure the microscopic EoS of DE. Therefore, a preference for $w(z)\neq -1$ in a non-interacting parameterization may be degenerate with a scenario in which dark matter and DE exchange energy and momentum. IDE models provide a natural framework for this possibility~\citep{Zhang:2005rg,Zhang:2005rj,Zhang:2007uh,Salvatelli:2013wra,Li:2015vla,Murgia:2016ccp,Pourtsidou:2016ico,Nunes:2016dlj,Wang:2016lxa,Kumar:2017dnp,DiValentino:2017iww,DiValentino:2019ffd,Lucca:2020zjb,DiValentino:2020vnx,Becker:2020hzj,Gao:2021xnk,Nunes:2021zzi,Yao:2022kub,Pan:2023mie,Zhai:2023yny,Escamilla:2023shf,Mishra:2023ueo,Forconi:2023hsj,Castello:2023zjr,Li:2023gtu,Giare:2024ytc,Li:2024qso,Halder:2024uao,Wang:2024vmw,Hoerning:2023hks,Silva:2025hxw,vanderWesthuizen:2025rip,Zhang:2025dwu,Li:2025muv,Yang:2025uyv,Wang:2025znm,Li:2025owk,Lyu:2025nsd,Pan:2025qwy,Wu:2025vrl,Wu:2024faw,Wu:2026qog}. 

In the standard $\Lambda$CDM model, CDM and DE are assumed to be separately conserved. The CDM density scales as $\rho_{\rm c}\propto a^{-3}$, while the energy density of the cosmological constant remains unchanged. However, the separate conservation of the two dark components is an assumption rather than a fundamental requirement. General covariance only requires the conservation of the total energy-momentum tensor. Thus, dark matter and DE may exchange energy and momentum through a non-gravitational interaction, provided that the total dark sector energy-momentum tensor remains conserved.

For the IDE model, the individual conservation equations for the energy-momentum tensors of DE and CDM are modified as
\begin{equation}
\label{eq:energyexchange}
\nabla_\nu T^{\nu}_{\mu,{\rm de}}
=
-\nabla_\nu T^{\nu}_{\mu,{\rm c}}
=
Q_\mu ,
\end{equation}
where $Q_\mu$ is the energy-momentum transfer four-vector between the two dark components. The opposite signs ensure that the total energy-momentum tensor of the dark sector remains covariantly conserved, while DE and CDM are not conserved separately. The choice of $Q_\mu$ specifies not only the background energy transfer rate $Q$, but also its perturbation $\delta Q$ and the momentum transfer rate $f$.

At the background level, the corresponding continuity equations can be written as
\begin{align}\label{conservation1}
{\rho}_{\rm de}^{\prime}+3\mathcal{H}(1+w)\rho_{\rm de}=aQ,\\
{\rho}_{\rm c}^{\prime}+3\mathcal{H}\rho_{\rm c}=-aQ,
\end{align}
where the prime denotes differentiation with respect to conformal time, $\rho_{\rm de}$ and $\rho_{\rm c}$ are the energy densities of DE and CDM respectively, and $\mathcal{H}=aH$ is the conformal Hubble parameter. The non-interacting case is recovered for $Q=0$.

In phenomenological IDE studies, $Q$ is usually assumed to be proportional to the energy density of CDM or DE~\citep{Amendola:1999qq,Billyard:2000bh}. Since $Q$ represents an energy-transfer rate, dimensional consistency requires the proportionality factor to contain a quantity with units of inverse time. A common and natural choice is the Hubble expansion rate $H$, leading to interaction forms such as
\begin{equation}
Q=\beta H\rho_{\rm c},\qquad Q=\beta H\rho_{\rm de},
\end{equation}
where $\beta$ is a dimensionless coupling parameter. These forms are simple and widely used, since the interaction rate scales with the cosmic expansion and the background conservation equations often remain analytically tractable. However, there is also a different viewpoint in the IDE literature. Since the interaction between dark matter and DE is local in nature, one may argue that the energy-transfer rate should not explicitly depend on the global expansion rate of the universe~\citep{Valiviita:2008iv,Boehmer:2008av,Caldera-Cabral:2008yyo,He:2008si,Clemson:2011an}. From this perspective, the Hubble constant $H_0$ can be introduced only as a fixed dimensional scale, leading to alternative phenomenological choices,
\begin{equation}
Q=\beta H_0\rho_{\rm c},\qquad Q=\beta H_0\rho_{\rm de}.
\end{equation}
These choices are observationally convenient and provide useful tests of dark-sector interactions, but they remain phenomenological parametrizations rather than unique consequences of an underlying microscopic theory.

A particularly simple and widely studied case is the interacting vacuum scenario, in which the intrinsic EoS of DE is fixed to $w=-1$ \citep{Li:2011ga,Xu:2011qv,Zhang:2012sya,Chimento:2013rya,Wang:2014xca,SanchezG:2014snn,Guo:2017hea,Guo:2017deu,Wang:2021kxc,Li:2024qso}. This choice represents a minimal extension of the base $\Lambda$CDM model, since it introduces an interaction in the dark sector without adding an evolving EoS parameter. However, in the presence of an interaction, setting $w=-1$ does not imply that the DE density is constant. From Eq.~(\ref{conservation1}), one obtains ${\rho}_{\rm de}^{\prime}=aQ$ for $w=-1$, so that $\rho_{\rm de}$ evolves whenever $Q\neq0$. This point can be made more explicit by rewriting the conservation equations in terms of effective EoSs,
\begin{align}\label{conservation2}
{\rho}_{\rm de}^{\prime}+3\mathcal{H}\left(1+w_{\rm de}^{\rm eff}\right)\rho_{\rm de}=0,\\
{\rho}_{\rm c}^{\prime}+3\mathcal{H}\left(1+w_{\rm c}^{\rm eff}\right)\rho_{\rm c}=0,
\end{align}
where
\begin{align}\label{effectiveEoS}
w_{\rm de}^{\rm eff}
=
w-\frac{aQ}{3\mathcal{H}\rho_{\rm de}},\\
w_{\rm c}^{\rm eff}
=
\frac{aQ}{3\mathcal{H}\rho_{\rm c}} .
\end{align}
Therefore, the effective EoS inferred from the background evolution is determined not only by the intrinsic value of $w$, but also by the form and sign of the interaction term $Q$. Even when $w=-1$, a non-zero interaction can lead to $w_{\rm de}^{\rm eff}\neq -1$, and may even mimic a time-varying or phantom-crossing EoS of DE in a non-interacting analysis.

A major theoretical issue in IDE is the stability of cosmological perturbations. Interactions modify not only the background evolution, but also the continuity and Euler equations for the dark components. In the usual fluid treatment, the DE pressure perturbation must be specified with care. IDE models can suffer from early-time large-scale instabilities: even for weak coupling and adiabatic initial conditions, curvature perturbations may diverge on super-Hubble scales~\citep{Valiviita:2008iv}. This result shows that background viability alone is not sufficient. Any IDE model intended to be confronted with CMB and large-scale-structure data must also possess a well-behaved perturbation sector.

To address this issue, \citet{Li:2014eha} developed the extended parametrized post-Friedmann framework for IDE, usually referred to as the ePPF framework\footnote{This framework builds upon the original parametrized post-Friedmann formalism \citep{Fang:2008sn}, which was initially introduced to address the DE perturbation issue when the EoS of DE crosses $-1$.}. The central idea is to avoid directly specifying the DE pressure perturbation on large scales, where the conventional fluid treatment may become ill defined. The ePPF approach introduces a parametrized relation between the momentum density of DE and that of the other components in the large-scale limit, and then consistently matches the large-scale and small-scale perturbation evolutions. This prescription removes the non-physical large-scale instability and yields well-behaved density and metric perturbations over a much broader IDE parameter space \citep{Li:2014eha,Zhang:2017ize}. Related large-scale stable IDE models have also been investigated~\citep{Li:2015vla}.

The development of \texttt{IDECAMB} further made this framework practically useful for observational cosmology~\citep{Li:2023fdk}. \texttt{IDECAMB} is an implementation of IDE cosmology in the Einstein--Boltzmann solver \texttt{CAMB}~\citep{Lewis:1999bs,Howlett:2012mh}. It provides a unified numerical interface for widely studied IDE models by introducing a parameterization with several mapping functions. By specifying these functions, both coupled fluid (CF) and coupled quintessence (CQ) scenarios can be implemented within the same computational framework. In particular, the perturbations of CF models are treated with the ePPF prescription in order to avoid possible large-scale instabilities. This is crucial in the DESI era, because one must test not only whether IDE can reproduce the DESI-preferred expansion history, but also whether it remains consistent with CMB anisotropies, CMB lensing and other structure-growth data.

In the post-DESI era, CF models provide one of the most direct phenomenological ways to test dark-sector interactions \citep{Giare:2024smz,Li:2024qso,Li:2025muv,Silva:2025hxw,Shah:2025ayl,Pan:2025qwy}. In these scenarios, dark matter and DE are treated as interacting cosmological fluids, with the interaction introduced directly through the modified conservation equations. The appeal of this approach is its flexibility: one can test whether simple energy-transfer terms can reproduce the DESI-motivated deviation from $\Lambda$CDM without requiring a time-varying intrinsic EoS of DE. \citet{Giare:2024smz} explored the implications of DESI DR1 BAO measurements for a specific IDE model with interaction kernel $Q=\xi H\rho_{\rm de}$, corresponding to an energy--momentum flow from dark matter to DE. Combining Planck CMB and DESI DR1 BAO data, they found a preference for a non-zero interaction at more than the $2\sigma$ confidence level and obtained a higher value of $H_0$, thereby reducing the tension with local distance-ladder measurements. Furthermore, \citet{Li:2024qso} compared four representative interacting-vacuum scenarios using Planck CMB, DESI DR1 BAO, and DESY5 SN data, demonstrating that the observational support for IDE depends sensitively on the assumed form of the interaction term $Q$. Their analysis shows that the preference for a non-zero coupling is mainly associated with models in which the energy-transfer rate is proportional to the DE density. In particular, the IDE model with $Q=\beta H_0\rho_{\rm de}$ exhibits an approximately $3\sigma$ departure from $\Lambda$CDM and leads to a significantly improved Akaike Information Criterion. This result suggests that the DESI-motivated deviation can be consistently interpreted within IDE frameworks, while also indicating that the combination of low- and high-redshift observations is beginning to distinguish which phenomenological interaction structures are more strongly supported by the current data.

A further issue is how the role of interaction changes once the EoS of DE is allowed to evolve. In this direction, \citet{Shah:2025ayl} reconsidered IDE models in light of DESI DR2, combining with Planck CMB and PantheonPlus data and considered interacting extensions of DDE parametrizations such as CPL and JBP. Their results show that the trends obtained with DESI DR2 are consistent with previous SDSS-based analyses, but with improved precision. These IDE models can alleviate the $S_8$ tension without exacerbating the $H_0$ tension. The EoS of DE exhibits an early phantom behaviour, in line with the DESI DR2 indication, and then evolves toward $w\simeq -1$ at lower redshifts. However, the statistical significance of excluding $w=-1$ is reduced once interaction is included.

\begin{figure}[t]
\begin{minipage}{0.4\textwidth}
\centering
\includegraphics[width=\linewidth]{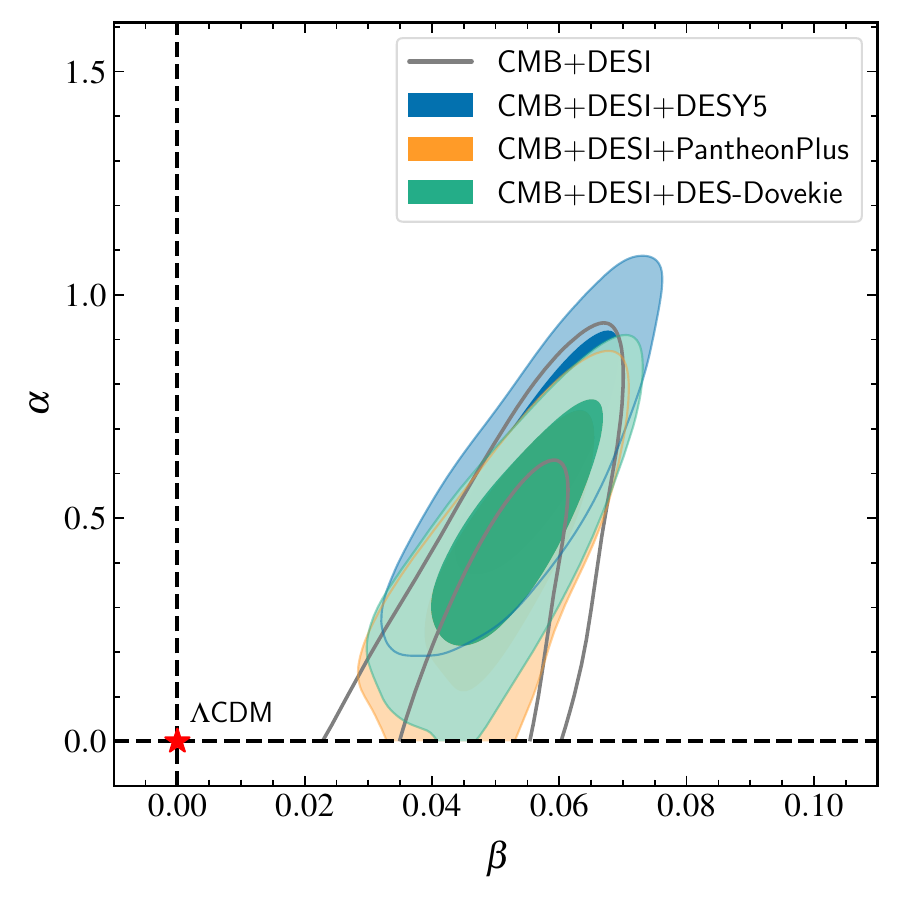}
\end{minipage}
\begin{minipage}{0.405\textwidth}
\centering
\includegraphics[width=\linewidth]{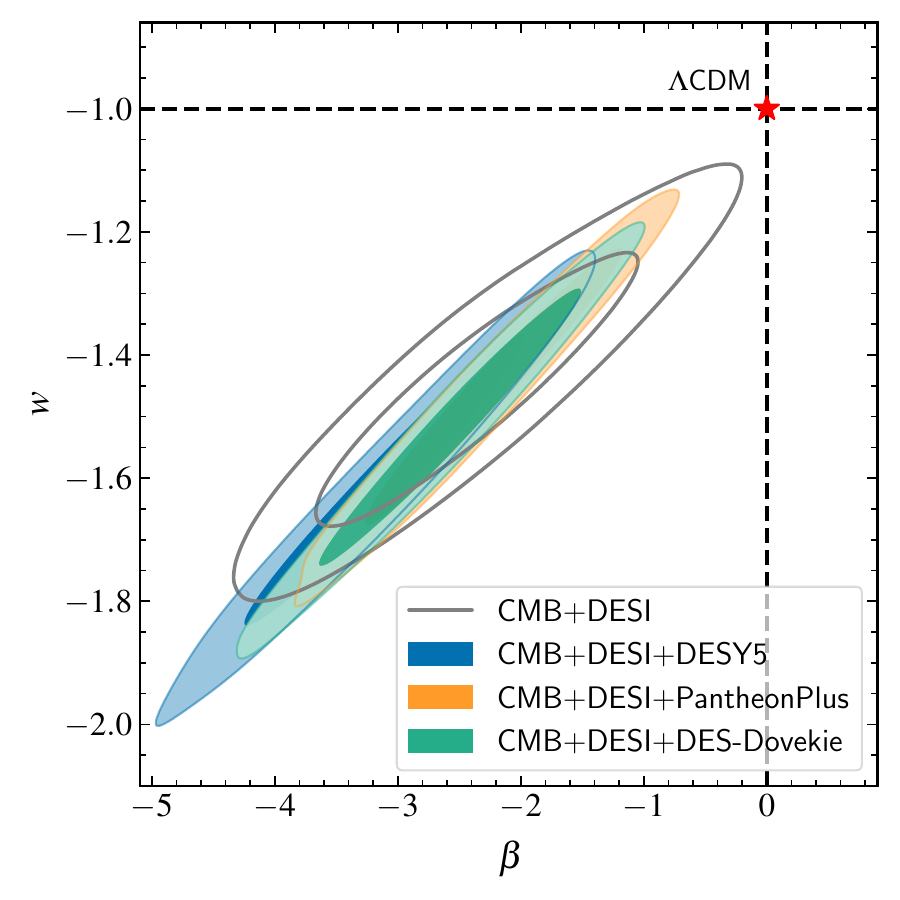}
\end{minipage}
\centering \caption{\label{fig4} Two-dimensional marginalized contours ($1\sigma$ and $2\sigma$ confidence levels) in the $\beta-\alpha$ plane within the CQ model (upper panel) and in the $\beta-w$ plane within the CF model (lower panel) from DESI, CMB, and SN data \citep{Li:2026xaz}.}
\end{figure}

The CF models discussed above are useful phenomenological probes, but the interaction term is usually specified directly at the level of the conservation equations. A complementary and more theory-oriented possibility is CQ, where DE is represented by a scalar field coupled to dark matter \citep{Amendola:1999er,Zhang:2005rg,Zhang:2005rj}. Unlike CF models, where $Q$ is imposed directly in the background equations, CQ implements the coupling at the level of the action or Lagrangian. The resulting effective energy transfer is then determined by the scalar-field dynamics and by the assumed coupling function. Although the coupling function itself remains phenomenological in most applications, this framework provides a more structured field-theoretic realization of dark-sector interactions and offers a useful framework between DESI-motivated deviations from $\Lambda$CDM and scalar-field models of late-time acceleration.

\begin{figure*}[htbp]
\resizebox{\textwidth}{!}{%
\includegraphics[]{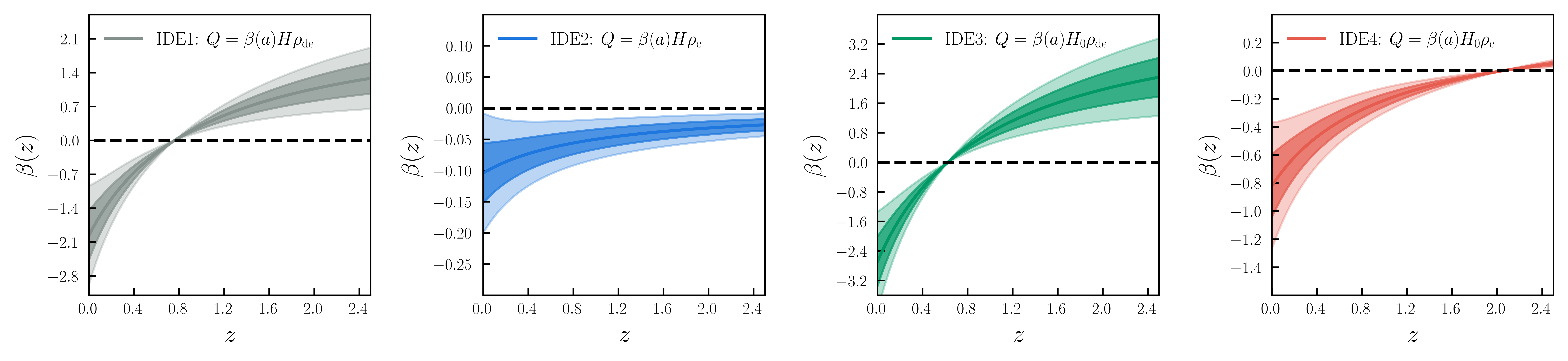}}
\centering
\caption{\label{fig5} Reconstructed evolutionary history of $\beta(z)$ at $1\sigma$ and $2\sigma$ confidence levels in the IDE1, IDE2, IDE3, and IDE4 models. The black dashed line in each plot represents the non-interacting line $\beta(z)=0$~\citep{Li:2025owk}.}
\end{figure*}

The CQ models have attracted renewed interest after DESI because they can produce an effective phantom-crossing behavior without requiring the scalar field itself to be phantom \citep{Antusch:2026ldp,Wang:2026vqw,Wang:2025znm}. A canonical quintessence field has a physical EoS satisfying $w_\phi>-1$. However, when the scalar field is coupled to dark matter, the effective EoS can cross the phantom divide. \citet{Chakraborty:2025syu} showed that a quintessence field with positive kinetic energy, coupled to dark matter through a Yukawa-type long-range interaction, can yield an observable effective EoS crossing $w_{\rm eff}=-1$ while keeping $w_\phi>-1$. For suitable parameter choices, the crossing occurs around $z\sim0.5$, with $w_{\rm eff}<-1$ in the past and $w_{\rm eff}>-1$ today, qualitatively reproducing the DESI-preferred CPL behavior. This provides a possible field-theoretic route to interpreting the DESI-motivated phantom crossing without introducing a fundamental phantom degree of freedom, although a full statistical exploration of the parameter space is still needed.

It is important to analyze CF and CQ realizations within a common computational framework. This is necessary because the same DESI-motivated departure from $\Lambda$CDM may correspond either to a phenomenological interaction between dark fluids or to a scalar-field realization of dark-sector coupling. Using the \texttt{IDECAMB} framework, \citet{Li:2026xaz} performed a unified analysis of both CQ and CF models with the latest CMB, DESI DR2 BAO, and SN data (including the DES-Dovekie recalibrated SNe). As shown in Figure~\ref{fig4}, the posterior distributions in the $\beta-\alpha$ plane (where $\alpha$ characterizes the steepness of the scalar field potential) for CQ and in the $\beta-w$ plane for CF consistently move away from the non-interacting $\Lambda$CDM limit. Across the considered data combinations, both realizations show evidence for non-vanishing dark-sector interactions at about the $3-5\sigma$ level. Furthermore, for the same number of free parameters, the CF and CQ models provide fits to the combined low- and high-redshift data that are at least comparable to, and in some cases better than, the CPL parameterization. Therefore, the DESI-motivated deviation should not be interpreted only as evidence for an evolving EoS. It can also be consistently accommodated within interacting dark-sector models, either in a CF description or in a scalar-field realization.

The significance of such unified analyses is conceptual as much as statistical. They show that IDE should not be treated merely as a secondary extension of DDE. Rather, it provides a physically distinct interpretation of the same background data. If a CF or CQ model can reproduce the DESI-preferred expansion history without requiring a time-varying intrinsic EoS of DE, then the central question changes: the issue is no longer only whether $w(a)$ evolves, but whether the assumption of separate conservation in the dark sector is valid. This shift in interpretation connects the DESI results to a broader question about the fundamental nature of the dark sector.

One interesting possibility is that the dark-sector interaction may itself be time dependent, and could even change sign during cosmic evolution. In most IDE models, the coupling parameter is assumed to be constant, so that the direction of energy transfer is fixed throughout cosmic history. While this assumption is convenient for phenomenological analyses, it is not imposed by fundamental physics. The possibility of a sign-changing dark-sector interaction was explored early by \citet{Cai:2009ht}. Rather than assuming a specific analytic form for the interaction, they adopted a piecewise parameterization in which the redshift range was divided into several bins and the coupling function $\delta(z)$, defined through $Q=3H\delta$, was treated as a constant in each bin. Their analysis suggested that the reconstructed interaction could cross the non-interacting line $\delta=0$, indicating a possible reversal in the direction of energy transfer. However, because the sign change was inferred from the best-fit values of several binned parameters, the conclusion was limited by the large observational uncertainties and by the difficulty of constraining multiple independent coupling parameters.

This limitation motivated a more economical parameterization proposed by \citet{Li:2011ga}, who introduced a running-coupling scenario in which the interaction term is written as $Q(a)=3\beta(a)H_0\rho_0$, where $\rho_0=\rho_{\rm de0}+\rho_{\rm c0}$ is the present-day total dark-sector density. In this framework, the time dependence of the interaction is encoded in a dimensionless coupling function,
\begin{equation}
\label{b}
\beta(a)=\beta_0 a+\beta_{\rm e}(1-a),
\end{equation}
with $\beta_0$ and $\beta_{\rm e}$ describing the late-time and early-time limits of the coupling, respectively. This two-parameter form provides a continuous interpolation between the early and late Universe, and allows the interaction to change sign during cosmic evolution if $\beta_0$ and $\beta_{\rm e}$ have opposite signs. Compared with a fully binned reconstruction, the running-coupling parameterization is more compact and therefore easier to constrain observationally.

\begin{figure*}[!htbp]
  \includegraphics[width=0.95\textwidth]{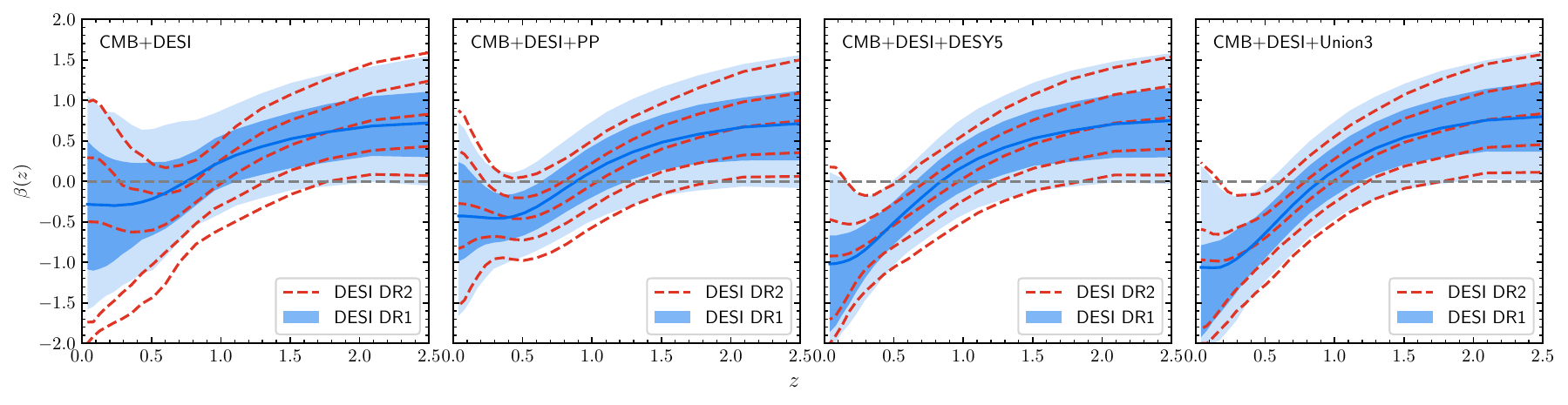}
  \centering
  \caption{\label{fig6} Reconstructed evolution of $\beta(z)$ with $1\sigma$ and $2\sigma$ confidence intervals for CMB+DESI and its combinations with individual SN datasets. Blue solid lines and shaded regions show DESI DR1-based reconstructions, red dashed lines show DESI DR2-based results, and black dashed lines denote the $\Lambda$CDM prediction \citep{Li:2025ula}.}
\end{figure*}

Recently, \citet{Li:2025owk} used the DESI DR2 BAO, together with DESY5 SN and Planck CMB data, to revisit sign-changeable IDE models within the running-coupling framework. They found that the coupling parameter tends to evolve from positive values at early times to negative values at late times, indicating a possible reversal in the direction of energy transfer during cosmic evolution, as illustrated in Figure~\ref{fig5}. In particular, the evidence for a sign-changing interaction can reach the $4.1\sigma$ level in the IDE3 model, and Bayesian evidence shows a moderate preference for the corresponding IDE scenarios over $\Lambda$CDM. These results suggest that, in the DESI era, allowing the interaction itself to vary with time may provide a viable extension of constant-coupling IDE models, although the conclusion remains dependent on the assumed interaction structure.

The DESI-era discussion of sign-changing interactions is not limited to the running-coupling reconstruction. Several recent studies have examined related possibilities in different IDE frameworks. \citet{Sabogal:2025mkp} proposed a sign-switching dark-sector coupling in which the direction of energy transfer reverses after the epoch when the dark matter and DE densities become comparable, and showed that this scenario may help alleviate both the $H_0$ and $S_8$ tensions, although the statistical preference depends on the SN calibration. \citet{Silva:2025hxw} compared conventional IDE models with sign-switching interactions and found that this scenario can lead to lower values of $S_8$, while the evidence for a non-zero coupling remains moderate. A broader class of time-dependent interactions was investigated by \citet{Yang:2025uyv}, who considered variable-coupling IDE models after DESI DR2 and found that evolving couplings can enhance the preference for interaction in some cases, although the conclusion is model dependent. Beyond the CF description, \citet{Wang:2026wrk} studied a non-minimally CQ model with an effective sign-switching interaction, providing a scalar-field realization of a changing energy-transfer direction. Overall, these studies suggest that sign-changing IDE models provide a promising direction for further investigation in light of DESI-era observations.

A particularly direct way to go beyond model-specific sign-changing IDE scenarios is to reconstruct the interaction history itself. In this direction, \citet{Li:2025ula} performed a non-parametric reconstruction of interacting vacuum energy, coupled to CDM, using DESI DR2 BAO, Planck CMB, and three SN compilations. Instead of assuming a prescribed analytic form for the coupling, they discretized the function $\beta(z)$ into 20 redshift bins and imposed a Gaussian smoothness prior. This strategy provides a less parameterization-dependent test of whether the dark-sector interaction undergoes a sign reversal during cosmic evolution.

As shown in Figure~\ref{fig6}, the reconstructed mean $\beta(z)$ changes sign, being positive at high redshifts and negative at late times. In the convention adopted in their work, this corresponds to energy transfer from CDM to DE in the earlier universe, followed by a reverse flow at late times. The high-redshift reconstruction shows an approximately $2\sigma$ deviation from the non-interacting limit, $\beta(z)=0$, while the low-redshift behavior depends on the SN sample. The strongest support is obtained for the CMB+DESI DR2+DESY5 combination, for which the reconstructed $\beta(z)$ model improves the goodness of fit by $\Delta\chi^2_{\rm MAP}=-17.76$ and yields Bayesian evidence $\ln\mathcal{B}=5.98\pm0.69$ relative to $\Lambda$CDM. Importantly, the improvement is not simply a consequence of introducing many unconstrained bins: their principal-component analysis shows that the data effectively constrain about three additional degrees of freedom. This result strengthens the view that the DESI-motivated departure from $\Lambda$CDM is not uniquely tied to an evolving EoS, but may also be reproduced by a sign-reversal interaction in the dark sector.

Overall, the DESI-motivated departure should not be interpreted solely as evidence for DDE. It may also point to the need to reconsider some of the basic assumptions about the dark sector in $\Lambda$CDM. In particular, at the level of background distances, DDE and IDE can produce similar expansion histories. Therefore, BAO and SN observations alone cannot uniquely determine whether the observed deviation originates from the intrinsic evolution of DE or from energy-momentum exchange between dark matter and DE. The key challenge is therefore to break this background-level degeneracy. While BAO and SNe mainly constrain the distance-redshift relation, distinguishing DDE from IDE requires perturbation-level information, such as the growth history of cosmic structure, lensing amplitudes, and redshift-space distortion measurements. In this sense, IDE provides not only an alternative interpretation of the DESI-motivated departure from $\Lambda$CDM, but also an important perspective for testing the standard assumption that dark matter and DE are separately conserved.

\subsection{Non-minimally coupled gravity}\label{subsec:nmc}

Model independent reconstructions of the EoS of DE from DESI data indicate a striking dynamical transition where $w$ crosses the phantom divide at low redshifts. This phantom crossing behavior has sparked widespread discussion regarding the necessity of extending the general relativity framework~\citep{Ye:2024ywg,Wolf:2024stt,BarrosoVarela:2024ozs,Wolf:2025jed,Liu:2025qca,Gao:2025onc,Pan:2025psn,Wang:2025znm}. From a theoretical perspective, standard canonical scalar field models such as minimally CQ are fundamentally incapable of achieving this transition because crossing the phantom divide typically induces a phantom instability. This pathology manifests as a wrong sign in the kinetic term of the scalar field action, which severely undermines the stability of the theory by rendering the total Hamiltonian unbounded from below at both the classical and quantum levels. 

To circumvent these instabilities within a mathematically consistent framework, \citet{Ye:2024ywg} utilized an effective field theory approach to systematically explore the broad landscape of Horndeski gravity. Their nonparametric reconstruction successfully isolated the non-minimal coupling operator $\Omega$ as the unique mechanism capable of stabilizing the effective DE fluid within the phantom domain. Consequently, they proposed the thawing gravity model, which is governed by the Lagrangian density
\begin{equation}
\mathcal{L} = \frac{M_{\text{Pl}}^2}{2}\left[1-\xi\left(\frac{\phi}{M_{\text{Pl}}}\right)^2\right]R + X - V_0 e^{-\lambda\phi/M_{\text{Pl}}}.
\end{equation}
Here, $M_{\text{Pl}}$ is the Planck mass, $R$ is the Ricci scalar, and $\xi$ is a dimensionless parameter magnifying the strength of the non-minimal coupling between the scalar field $\phi$ and gravity. $X$ represents the canonical kinetic term, while $V_0$ and $\lambda$ govern the energy scale and slope of the exponential potential, respectively. Their result demonstrates that a non-minimally coupled scalar field can naturally accommodate a safe phantom crossing while providing an improved fit to the joint DESI DR1 BAO, CMB, and SN data. Subsequent refinements using extended parameterizations of the effective field theory functions have further substantiated these results, yielding a robust detection of non-minimal coupling at a $\sim3\sigma$ deviation \citep{Pan:2025psn}. These studies demonstrate that a flexible second order Taylor expansion of the coupling function with respect to the DE fraction is required to accurately capture the dynamic feature reconstructed near $z \approx 0.6$. 

\begin{figure}[htbp]
\centering
\includegraphics[width=0.48\textwidth]{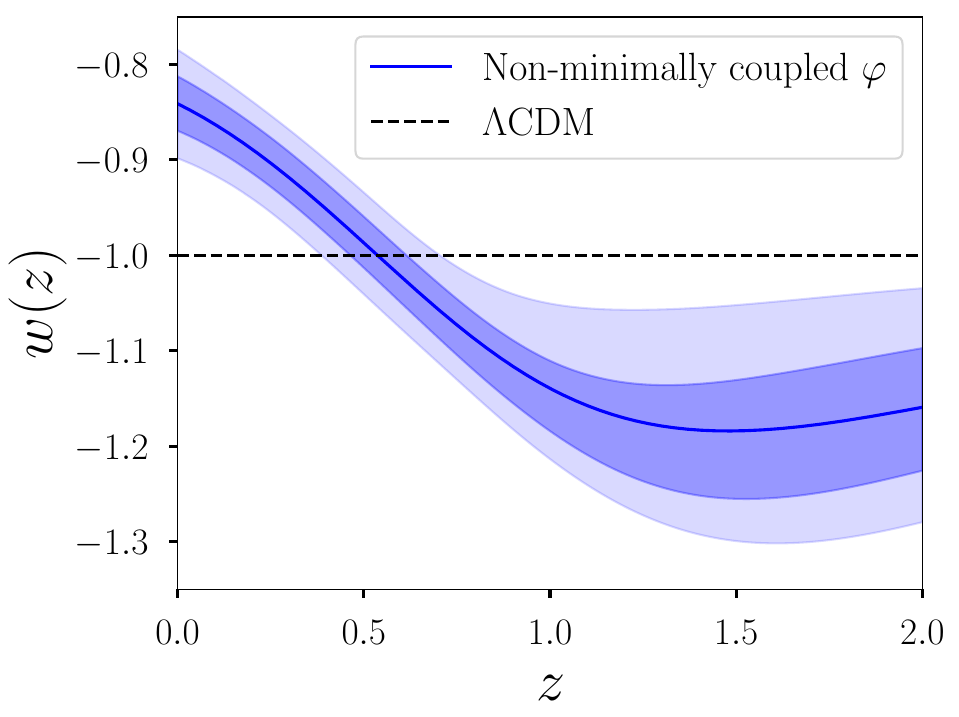}
\caption{Reconstruction of the EoS of DE for the non-minimally coupled scalar field model. The shaded regions denote the confidence intervals, illustrating that the EoS remains in the phantom regime at higher redshifts before sharply thawing and crossing the boundary at $z \approx 0.5$ \citep{Wolf:2025jed}.}
\label{fig7}
\end{figure}

Furthermore, to fully realize the rapid late-time dynamical evolution preferred by DESI, \citet{Wolf:2024stt} introduced a generalized scalar field Lagrangian density to capture a broader class of scalar-tensor theories,
\begin{equation}
\mathcal{L} = \frac{M_{\text{Pl}}^2}{2} F(\varphi) R - \frac{1}{2} G(\varphi) X - V(\varphi) - J(\varphi) X^2,
\end{equation}
where $F(\varphi)$ dictates the non-minimal coupling to gravity, $G(\varphi)$ and $J(\varphi)$ modify the kinetic dynamics, and $V(\varphi)$ is the scalar potential. Focusing on the thawing regime ($G=1$, $J=0$), the non-minimal coupling and the potential can be expanded to quadratic order as
\begin{align}
F(\varphi) &\simeq 1 - \xi \frac{\varphi^2}{M_{\text{Pl}}^2}, \\
V(\varphi) &\simeq V_0 + \beta\varphi + \frac{1}{2}m^2\varphi^2.
\end{align}
Here, $\xi$ determines the coupling strength, $V_0$ is the baseline potential energy, $\beta$ is the initial linear slope, and $m^2$ represents the effective mass squared, governing the curvature of the potential. Specifically, they proposed the hilltop thawing gravity (HTG) model, which features a concave-down potential satisfying $m^2 \equiv d^2V/d\varphi^2 < 0$. Unlike standard quadratic or exponential models, a concave hilltop feature causes the scalar field to roll extremely rapidly once it begins moving. This rapid acceleration translates into a sharp, nonlinear thawing evolution in the EoS, providing the necessary mathematical freedom to fully cover the $(w_0, w_a)$ parameter space favored by DESI. 

Building on this theoretical foundation, \citet{Wolf:2025jed} comprehensively evaluated the Bayesian evidence for HTG models. Their analysis revealed that standard minimally coupled scalar fields ($\xi=0$) are strongly disfavored by the latest observations because they cannot replicate the rapid late time variation in the expansion rate preferred by the DESI data. In contrast, incorporating the non-minimal coupling term $\xi \varphi^2 R$ allows the scalar field to remain potential energy dominated early on and subsequently driven into a past phantom regime during the matter dominated era. This mechanism yields an exceptionally strong statistical preference over the $\Lambda$CDM model, achieving a remarkable Bayes factor of $\log(B) = 7.34 \pm 0.6$ when evaluated by Planck CMB, DESI DR2 BAO, and DESY5 SN data. The resulting evolution of the reconstructed EoS of DE is depicted in Figure \ref{fig7}. The trajectory explicitly reveals that the EoS resides within the phantom regime across intermediate redshifts before undergoing a transition to cross the phantom divide at $z \approx 0.5$ and reach $w_0 > -1$ today. However, despite these phenomenological successes, it remains crucial to emphasize that such extensive modifications to the gravitational action inevitably imply a time-varying effective gravitational constant on cosmological scales. Therefore, the viability of this theory strictly relies on the implementation of efficient small-scale screening mechanisms to satisfy the tight local experimental bounds on fifth forces.

Overall, the dynamic transition into the phantom regime suggested by DESI can be theoretically accommodated without pathological instabilities by invoking a non-minimal coupling between the scalar field and the gravitational curvature. When combined with a concave hilltop potential, this framework provides a physically consistent and statistically favored description of cosmic acceleration, underscoring the necessity of exploring extended gravitational theories in the era of high-precision cosmology.

\subsection{Non-standard dark matter}\label{subsec:ncdm}

An alternative approach to interpreting the discrepancies highlighted by the DESI data involves relaxing the standard assumption that dark matter is a perfectly cold and pressureless fluid. Although the standard CDM approximation proves highly successful on cosmological scales, it lacks a rigorous derivation from fundamental first principles. Consequently, various studies have investigated whether a non-zero dark matter EoS could partially account for the observed late time background signals~\citep{Kumar:2025etf,Li:2025eqh,Li:2025dwz,Wang:2025zri}.

\begin{figure}[htbp]
\centering
\includegraphics[width=0.49\textwidth]{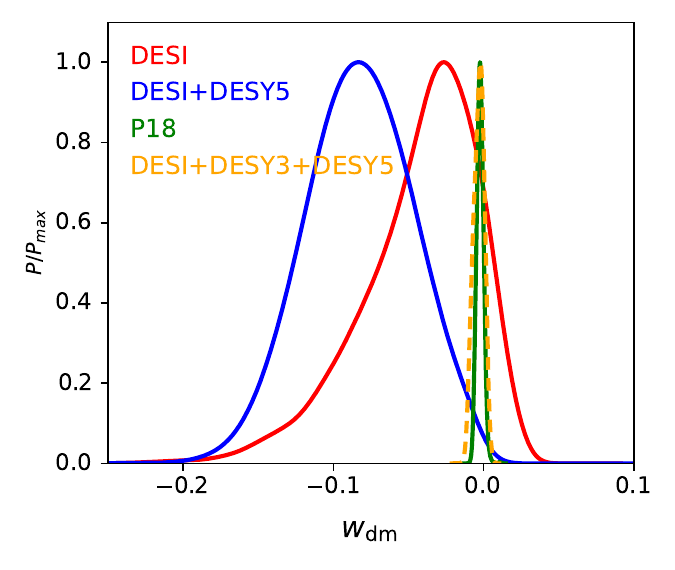}
\caption{The one-dimensional marginalized posterior distributions of the dark matter EoS parameter $w_{\rm dm}$ for various dataset combinations \citep{Kumar:2025etf}.}
\label{fig8}
\end{figure}

A fundamental realization of this concept is the constant EoS model investigated by \citet{Kumar:2025etf}. As shown in Figure~\ref{fig8}, utilizing DESI DR2 BAO, DESY5 SNe, Planck data, and DESY3 weak lensing information, their analysis revealed that DESI data alone mildly prefer a non-zero $w_{\rm dm}$. Furthermore, the combination of DESI and DESY5 data yields $w_{\rm dm}=-0.084\pm0.035$, thereby excluding the standard CDM paradigm at a significance of approximately $2.4\sigma$ \citep{Kumar:2025etf}. However, the inferred sign and magnitude of the dark matter EoS exhibit strong dependencies on the specific dataset combinations employed. Specifically, the combination of Planck and DESI data suggests a marginally positive value, whereas the inclusion of weak lensing measurements tends to shift the constraints back towards the standard pressureless scenario. This variance complicates the physical interpretation. The observed deviation does not constitute an unambiguous detection of non-standard dark matter properties. Rather, it indicates that the dark matter sector possesses the capacity to absorb portions of the cosmological tension when the DE component is held fixed.

This dependence of dataset combinations is further elucidated in the study by \citet{Li:2025eqh}, who analyzed a non-zero dark matter EoS within the phenomenologically emergent DE framework. They found that the combination of CMB, DESI DR2 BAO, and DESY5 prefers a negative $w_{\rm dm}$ at around the $3\sigma$ level. However, this significance is reduced to approximately $2\sigma$ when substituting the PantheonPlus sample, and the overall Bayesian evidence continues to favor the standard $\Lambda$CDM model over this extended scenario \citep{Li:2025eqh}. Furthermore, the positive correlation between the dark matter EoS parameter and the Hubble constant in this model implies that introducing a negative $w_{\rm dm}$ fails to further alleviate the $H_0$ tension.

A compelling demonstration of the parameter degeneracy between the dark matter and DE sectors is provided by \citet{Li:2025dwz}. Their analysis considered a non-zero $w_{\rm dm}$ across three distinct forms for the EoS of DE, including the cosmological constant, a constant EoS, and the CPL parameterization. When DE is fixed to the cosmological constant or a constant EoS, the combination of CMB, DESI DR2 BAO, and SN prefers a non-zero $w_{\rm dm}$ at a significance between $2.8\sigma$ and $3.3\sigma$. Conversely, when the CPL parameterization is introduced, this significance diminishes to between $0.8\sigma$ and $1.1\sigma$ \citep{Li:2025dwz}. In other words, the statistical preference for non-standard dark matter diminishes considerably when DE is permitted to evolve dynamically. This clearly demonstrates that the evidence for non-standard dark matter depends on the assumptions of the DE model.

A further extension is the dynamical dark matter scenario proposed by \citet{Wang:2025zri}, wherein the dark matter EoS inherently evolves over cosmic time. In their framework, individual datasets and some combinations produce approximately $2\sigma$ evidence for dark matter evolution, allowing for the coexistence of dynamical dark matter and DDE. Although certain speculative implications regarding the ultimate fate of the universe require cautious interpretation, the fundamental conclusion remains significant. Relaxing the assumption of a pressureless dark matter fluid significantly increases the model dependence of any subsequent DE inferences \citep{Wang:2025zri}.

Overall, current DESI observations provide a degree of statistical evidence for the existence of non-standard dark matter, suggesting that the deviations revealed by the DESI data can alternatively be accommodated by non-standard dark matter hypotheses. However, the introduction of such non-standard assumptions can significantly impact the inferred evidence for DDE, primarily due to the strong underlying parameter degeneracies between the two dark sectors. Consequently, future high-precision observations and independent cosmological probes will be essential to break these degeneracies and rigorously test these competing theoretical frameworks.

\section{Impact on current cosmological tensions}\label{sec:tensions}

Recent DESI observations have driven a critical re-evaluation of the principal cosmological tensions, namely the $H_0$ tension, $S_8$ tension, and neutrino mass problem. While DDE can loosen constraints on neutrino mass and naturally ease tensions with particle physics experiments, it often falls short in resolving other discrepancies like the $H_0$ tension when anchored to early-universe constraints. This section discusses how these ongoing tensions act as robust discriminants for new physics, exploring how early-universe extensions, modified gravity, and IDE offer complementary pathways for reconciling the $H_0$, $S_8$, and $\sum m_\nu$ anomalies within a unified theoretical framework.

\subsection{The $H_0$ tension}

The $H_0$ tension remains one of the most statistically significant observational discrepancies in modern cosmology \citep{Guo:2018ans,Verde:2019ivm,Planck:2018vyg,Vagnozzi:2019ezj,DiValentino:2020zio,DiValentino:2021izs,Shah:2021onj,Vagnozzi:2021gjh,Gao:2021xnk,Riess:2021jrx,Perivolaropoulos:2021jda,Schoneberg:2021qvd,Abdalla:2022yfr,DiValentino:2022fjm,Kamionkowski:2022pkx,Giare:2023xoc,Hu:2023jqc,Vagnozzi:2023nrq,Song:2022siz,Jin:2023sfc,Gao:2022ahg,Zhang:2024rra,CosmoVerseNetwork:2025alb,Leauthaud:2025azz,Song:2025ddm,Jin:2025dvf,DiValentino:2026uua}. The recent data release from DESI further complicates this landscape by hinting at DDE. Specifically, the data show a preference for an evolving EoS that exhibits a phantom crossing behavior, which fails to alleviate the $H_0$ tension because fitting this DDE with CMB data generally infers a lower present-day expansion rate. Consequently, the $H_0$ tension has been widely discussed in the post-DESI era, prompting comprehensive reviews and broad theoretical evaluations. Theoretical efforts aim to alleviate the tension by exploring late-time phenomenological modifications \citep{Li:2019yem,Li:2020ybr,Li:2026hwq}, interacting dark sectors \citep{DiValentino:2019ffd,Gao:2021xnk,Giare:2024smz}, early universe physics \citep{Poulin:2018cxd,Ye:2020btb,Yin:2023srb}, modified gravity theories \citep{Hogas:2025ahb,Kavya:2025vsj,Wang:2025znm}, as well as local environmental effects and model-independent systematics \citep{Teixeira:2025czm,Liu:2025evk,Du:2025csv,Zhou:2025rvf,Pantos:2026rpe,Pedrotti:2026dwj}.

Despite the observational hints from DESI, pure late-time DDE faces theoretical difficulties in resolving the $H_0$ tension \citep{Jia:2025poj,Zhou:2025kws,Pedrotti:2025ccw,Zhang:2025lam,Li:2025vqt,Xu:2026sbw,Bansal:2026axl}. \citet{Pedrotti:2025ccw} demonstrate that unanchored SN data place stringent geometric constraints on the low-redshift expansion history. Their analysis shows that purely late-time solutions fail to provide a high $H_0$ value because boosting the expansion rate disrupts the fit to SN relative distances. Conversely, \citet{Jia:2025poj} use a non-parametric reconstruction to suggest that the tension might reflect an evolving expansion rate, a perspective also explored in various phenomenological and holographic DE models \citep{Adi:2025hyj,Li:2025vqt}. They find that an evolving EoS naturally yields a decreasing $H_0$ toward higher redshifts, which can partially bridge the gap between local measurements and early universe constraints.

IDE models offer a phenomenological alternative to simple late-time dynamical models. These models can effectively simulate the background dynamic behavior of DDE while introducing distinct imprints at the perturbation level \citep{Giare:2024smz,Zhang:2025dwu,Smith:2025uaq,Buen-Abad:2025bgd,Yashiki:2025loj,Das:2025asx,Garny:2026ish,Schiavone:2026agq}. \citet{Giare:2024smz} investigate a scenario with an energy-momentum flow from dark matter to DE. Their results show that this coupling mechanism yields a higher present-day expansion rate of approximately $H_0 \sim 71.5\,\rm km\,s^{-1}\,Mpc^{-1}$, thereby reducing the discrepancy with local distance ladder measurements to around $1.5\sigma$ level. Additionally, \citet{Zhang:2025dwu} explore a new IDE model which provides a field-theoretical basis for the interaction and further demonstrates the capability of coupled dark sectors to ease the $H_0$ tension.

Early dark energy (EDE) fundamentally reduces the sound horizon prior to recombination to address the $H_0$ tension \citep{Poulin:2018cxd,Sakstein:2019fmf,Ye:2020btb}. Combining ACT and DESI DR2 BAO data, \citet{Poulin:2025nfb} yield a $2.5\sigma$ statistical preference for EDE and successfully lowers the $H_0$ tension to below $2\sigma$. Moreover, \citet{Bella:2026zuk} propose a multi-field EDE framework to overcome the stringent polarization constraints from high-multipole CMB data that challenge standard single-field models. By utilizing double axion fields, this model smoothly distributes the energy injection over a broader redshift window and achieves a peak fractional EDE contribution of approximately ten percent. Their analysis reveals that this extended injection effectively breaks the high-multipole bottleneck and can yield a higher Hubble constant value of around $H_0 \sim 72\,\rm km\,s^{-1}\,Mpc^{-1}$. However, with CPL-like DE this pre-recombination resolution of $H_0$ tension could suppress the preference of DESI for DDE~\citep{Pang:2024wul,Wang:2025djw,Pang:2025lvh,Wang:2024dka}. Thus it is necessary to be cautious about the evidence of DDE in a potential $H_0$ tension-free cosmology. See the references \citep{Seto:2024cgo,Wang:2024tjd,Poulin:2024ken,Qu:2024lpx,SPT-3G:2025vyw,Yashiki:2025loj,Chaussidon:2025npr,Jiang:2025hco,Peng:2025tqt,Reeves:2025xau,Toda:2025kcq,Jhaveri:2026bla,Yin:2026gss} for other analysis of EDE with DESI data.

\begin{figure}[htbp]
\centering
\includegraphics[width=0.48\textwidth]{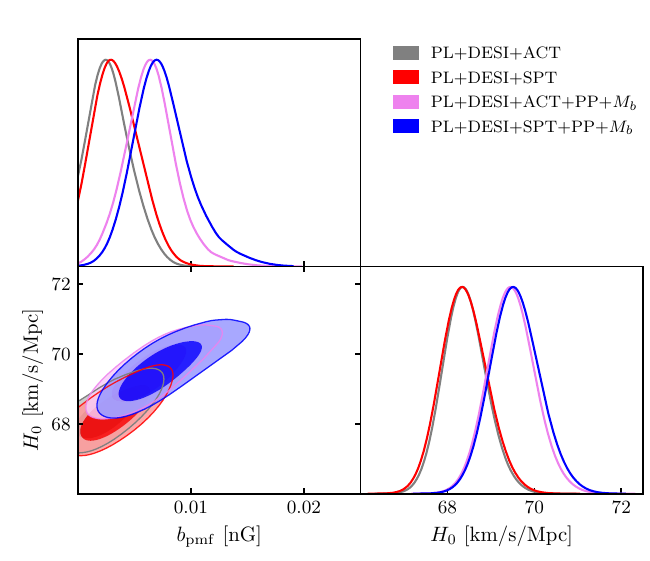}
\caption{Marginalized joint $1\sigma$ and $2\sigma$ confidence level contours for $H_0$ and the PMFs parameter $b_{\rm pmf}$ for various data combinations \citep{Jedamzik:2025cax}.}
\label{fig9}
\end{figure}

Modifications to the early universe and the standard recombination history present alternative physical solutions \citep{B:2025koi,Li:2025nnk,Chluba:2025jqt,Jedamzik:2025cax,Sailer:2025lxj,Carloni:2025jlk}. \citet{Jedamzik:2025cax} demonstrate that primordial magnetic fields (PMFs) can alter the recombination process. As shown in Figure~\ref{fig9}, they find that small-scale baryon clumping induced by $5$ to $10$ pico-Gauss PMFs accelerates the recombination epoch and successfully yields $H_0 \sim 71\,\rm km\,s^{-1}\,Mpc^{-1}$ without conflicting with high-multipole CMB data. Complementary to this, \citet{Sailer:2025lxj} investigate the reionization optical depth, an approach supported by related studies on cosmic birefringence and extra radiation \citep{Lee:2025yah,Huang:2025xyf,Allali:2025yvp,Wang:2025dzn}. They find that assuming a higher reionization optical depth near $\tau \sim 0.09$ intrinsically raises the expansion rate to $H_0 \sim 67.9\,\rm km\,s^{-1}\,Mpc^{-1}$ and suppresses the statistical preference for DDE.

Modified gravity theories provide a theoretical framework for reconciling observational discrepancies by geometrically altering the expansion dynamics \citep{Kavya:2025vsj,Efstratiou:2025iqi,Plaza:2025nip,Wang:2025znm,Montani:2025nmz,Legner:2025hrt,Najera:2025htf,Bouhmadi-Lopez:2026dte}. \citet{Hogas:2025ahb} show that bimetric gravity naturally produces phantom DE behavior governed by fundamental physical constraints. Their analysis indicates that this model yields $H_0 \sim 69.0\,\rm km\,s^{-1}\,Mpc^{-1}$ and eases the tension from over $5\sigma$ to $3.7\sigma$. Furthermore, \citet{Najera:2025htf} investigate symmetric teleparallel gravity utilizing a specific exponential modification to the non-metricity scalar. Their statistical analysis indicates that this exponential extension dynamically increases the late-time expansion rate and harmonizes the DESI data with local calibrations. Both approaches avoid the introduction of phenomenological fluids and maintain the constrained early universe acoustic horizon.

It remains essential to investigate whether the $H_0$ tension stems from localized environmental effects \citep{Giani:2024nnv,Teixeira:2025czm,Liu:2025evk,Ong:2026tta,Pantos:2026cxv}. \citet{Zhou:2025rvf} utilize multiple independent tracers across early and late epochs to confirm the macroscopic isotropy of the Hubble expansion, which rules out spatial directional anomalies. Conversely, local environmental variations such as the local supervoid model remain actively discussed \citep{Stiskalek:2025ibp}. \citet{Banik:2026imu} show that if the observer resides within a massive underdense region, the resulting outward peculiar velocities and gravitational redshifts naturally induce a higher localized expansion rate. This mechanism offers an explanation for the low-redshift distance anomalies independent of background DDE.

Overall, within the context of DESI, the $H_0$ tension remains a formidable cosmological challenge that cannot be straightforwardly resolved by simple late-time DDE models. Various theoretical solutions, including IDE, early universe modifications (such as EDE and PMFs), modified gravity, and local environmental effects, offer viable pathways to alleviate this tension. Ultimately, determining whether the $H_0$ tension points toward fundamental new physics in the early universe, alternative gravitational dynamics, or localized systematic effects will require future high-precision, independent cosmological probes.

\subsection{The $S_8$ tension}

The $S_8$ tension, which characterizes the discrepancy in the amplitude of matter density fluctuations between early universe CMB predictions and large-scale structure probes, remains an unresolved issue in standard cosmology \citep{DiValentino:2020vvd,CosmoVerseNetwork:2025alb,Pantos:2026koc,DiValentino:2026uua}. Recent evaluations reveal significant variations across different observational probes. As noted by \citet{DiValentino:2026uua}, while surveys like DESY6~\citep{DES:2026fyc,DES:2026mkc} exhibit tension with the CMB baseline~\citep{Planck:2018vyg}, others such as KiDS-Legacy~\citep{Wright:2025xka,Reischke:2025hrt} are consistent with it. This measurement divergence indicates that survey specific systematic errors, including photometric redshift calibration, intrinsic alignments, and complex baryon feedback, are potential sources of the current $S_8$ tension, although new physics cannot be entirely ruled out.

Before invoking new physics, conventional extensions within the standard framework have been tested. \citet{Terasawa:2025fpf} demonstrate that incorporating parameterized baryon feedback or freely varying massive neutrinos does not yield a statistically preferred fit to combined cosmological data. Phenomenological modifications at the perturbation level present alternative results. \citet{Giare:2025ath} report that introducing a suppressed structure growth index ($\gamma > 0.55$) lowers $S_8$ to ease this tension and relaxes the upper bounds on neutrino masses without altering the $\Lambda$CDM background evolution.

Modified gravity theories offer a mechanism to dynamically alter late time structure growth~\citep{Terasawa:2025fpf,Li:2025msm,Kavya:2025vsj,Du:2026cly}. \citet{Terasawa:2025fpf} find a statistical preference for phenomenological models where the growth suppression occurs during the DE dominated epoch. In the context of symmetric teleparallel $f(Q)$ gravity, \citet{Li:2025msm} introduce a square root correction that decouples the perturbation evolution from the background expansion. This approach dampens the late time clustering rate and reduces the discrepancy with redshift space distortion data. Similarly, \citet{Kavya:2025vsj} demonstrate that a power law $f(Q)$ model yields a lower $S_8$ and a higher $H_0$, providing a geometric explanation for multiple cosmological tensions.

\begin{figure}[htbp]
\centering
\includegraphics[width=0.49\textwidth]{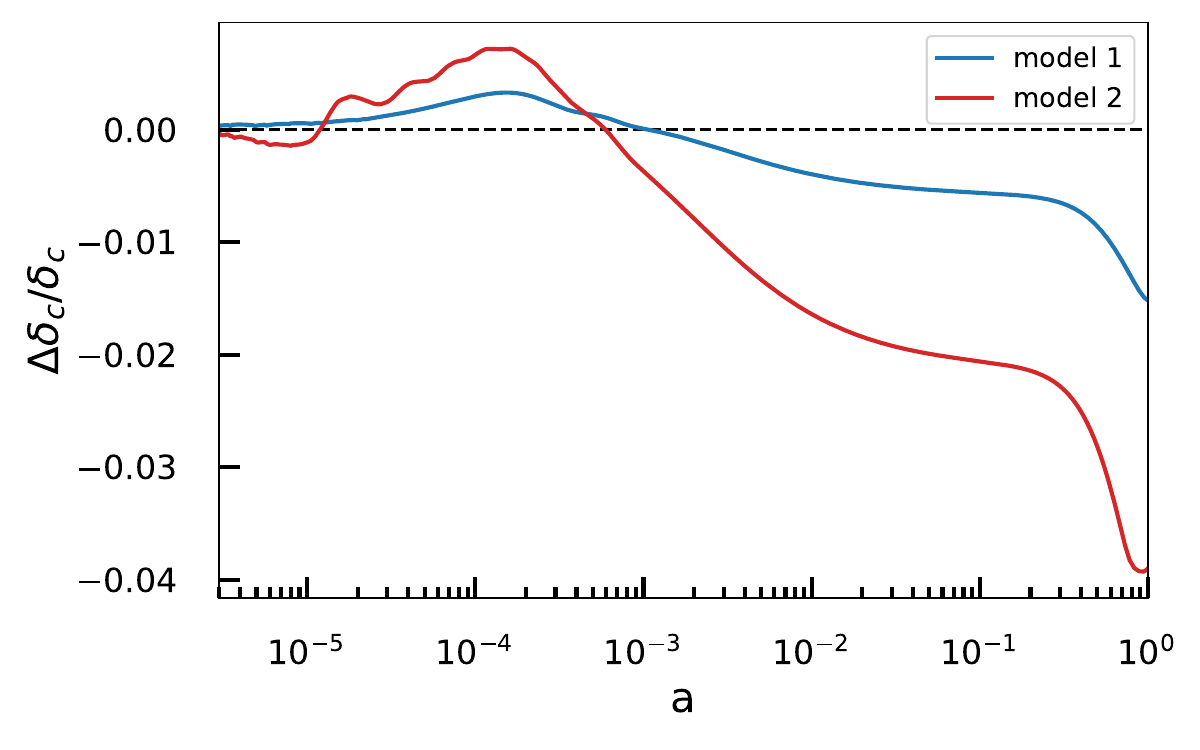}
\caption{The fractional difference in CDM density perturbations, $\Delta\delta_c/\delta_c$, as a function of the scale factor $a$ for the dark axion and dark baryons interaction model \citep{Khoury:2025txd}. Here, ``model 1'' and ``model 2'' correspond to two fiducial models with distinct parameter choices for the dark sector coupling ($\sigma_N/m_0$, where $\sigma_N$ is the pion-nucleon sigma term and $m_0$ is the dark baryon mass), axion decay constant ($f$), and vacuum energy parameter ($v$). Specifically, Model 1 assumes $\sigma_N/m_0 = 0.01$, $f = 0.25 M_{\rm Pl}$, and $v = 0$, whereas Model 2 adopts $\sigma_N/m_0 = 0.02$, $f = 0.035 M_{\rm Pl}$, and $v = 4$. The curves illustrate the late time suppression of perturbation growth.}
\label{fig10}
\end{figure}

Interacting dark sectors provide another approach to modify structure formation. \citet{Khoury:2025txd} propose an interaction between dark axions and dark baryons. As shown in Figure~\ref{fig10}, this mechanism delays the matter radiation equality epoch and suppresses the late time growth of matter perturbations ($\Delta\delta_c/\delta_c < 0$), which lowers $S_8$ by $1.5\%$ to $4\%$. Furthermore, because standard EDE models typically increase the $S_8$ tension by raising the CDM density \citep{Pang:2025lvh,Smith:2025uaq}, \citet{Yashiki:2025loj} construct a hybrid model combining EDE with an interacting dark sector. The energy transfer from DE to dark matter in this framework suppresses the small scale power spectrum, aiming to address both the $H_0$ and $S_8$ tensions. Furthermore, \citet{Wang:2025znm} show that non-minimally CQ can alter the effective dark matter evolution to alleviate the $S_8$ tension.

In addition to interactions purely within the dark sector, non-standard couplings between neutrinos and dark matter have emerged as a highly compelling resolution to the $S_8$ tension. Recently, \citet{Zu:2025lrk} demonstrated that a non-standard scattering between neutrinos and dark matter modifies the energy and momentum exchange in the early universe. This interaction induces dark acoustic oscillations subject to diffusion damping, which effectively suppresses structure growth on small scales. By incorporating nonlinear corrections from N-body simulations into a joint analysis of ACT high-multipole data, Planck CMB, DESI DR2 BAO, and DESY3 cosmic shear measurements, they found that an interaction strength of $u_{\nu \rm DM} \sim 10^{-4}$ harmonizes the $S_8$ values across these different observational epochs. Notably, the ACT high-$\ell$ data independently exhibit a preference for this non-zero interaction, which is further bolstered by DESY3, achieving a combined statistical significance approaching $3\sigma$. Compared to warm dark matter scenarios, this $\nu$-dark matter interaction offers greater flexibility in evading stringent CMB bounds.

Modifications to early universe physics and primordial conditions can also impact the $S_8$ parameter. \citet{Wang:2025dzn} introduce a varying electron mass combined with a free CMB lensing amplitude, which pulls $S_8$ down to approximately $0.808$ while lowering the $H_0$ tension. Alternatively, \citet{Li:2025nnk} explore the replacement of the standard power law spectrum with the Harrison Zel'dovich (HZ) primordial power spectrum ($n_s=1$). While the HZ spectrum lowers $S_8$ to $0.7645$, it increases the intrinsic CMB lensing and spatial curvature anomalies, indicating that such primordial adjustments face observational constraints. Additionally, variations in the reionization optical depth ($\tau$) have been shown to be statistically coupled with the $S_8$ parameter, underscoring the degeneracy between early universe modeling and late time clustering constraints \citep{Allali:2025yvp}.

Overall, while the $S_8$ tension may partly stem from survey-specific systematic errors, it remains a powerful discriminant for probing new physics. A wide array of theoretical frameworks, ranging from late-time modified gravity and dark sector interactions to early-universe scattering and primordial adjustments, demonstrates that suppressing structure growth offers viable and diverse pathways to reconcile this cosmological tension.

\subsection{The neutrino-mass problem}\label{subsec:nu}

Neutrino oscillation experiments have provided compelling evidence that neutrinos possess non-zero masses, constituting the only experimentally established manifestation of physics beyond the Standard Model to date. By measuring the mass-squared differences among the three active neutrino mass eigenstates, these experiments establish a lower bound on the total neutrino mass of $\sum m_\nu \gtrsim 0.06\,{\rm eV}$ for the normal hierarchy (NH) and $\sum m_\nu \gtrsim 0.10\,{\rm eV}$ for the inverted hierarchy \citep{Esteban:2020cvm,deSalas:2020pgw,Esteban:2024eli,Denton:2025jkt}. Meanwhile, laboratory kinematic probes via $\beta$-decay, such as the KATRIN experiment, place an upper bound of $\sum m_\nu < 1.35\,{\rm eV}$ \citep{KATRIN:2024cdt}. 

On the other hand, cosmological observations provide a highly sensitive and complementary avenue to constrain the absolute neutrino mass. Analyses combining Planck CMB data with SDSS BAO yield an upper limit of $\sum m_\nu < 0.12\,{\rm eV}$~\citep{Planck:2018vyg}. Recently, combining Planck CMB data with the DESI DR2 BAO data within the standard $\Lambda$CDM model has further tightened this cosmological upper limit to $\sum m_\nu < 0.064\,{\rm eV}$ \citep{DESI:2025zgx}. This stringent bound already approaches the physical lower limit of the NH, thereby creating a notable tension between cosmological inferences and terrestrial laboratory constraints. 

Consequently, exploring physical mechanisms to alleviate this neutrino-mass problem under the context of DESI observations has emerged as a highly active research direction, encompassing comprehensive parameter constraints and mass hierarchy evaluations \citep{DiValentino:2024xsv,Loverde:2024nfi,Wang:2024hen,Du:2024pai,Shao:2024mag,Escudero:2024uea,Herold:2024nvk,Elbers:2025vlz,Giare:2025ath,Zhou:2025nkb,Reischke:2025hrt,Ivanov:2026dvl,Feng:2026pzs}, investigations into statistical systematics and physical interpretations of negative mass preferences \citep{Yeung:2024krv,Green:2024xbb,Elbers:2025vlz,Jhaveri:2025neg,RoyChoudhury:2025dhe,Namikawa:2025doa,Graham:2025dqn,Chebat:2025kes,Sharma:2025iux,Pulido-Hernandez:2026hcs}, explorations of DDE and IDE models \citep{Wang:2024hen,Allali:2024aiv,Jiang:2024viw,Du:2024pai,Yadav:2024duq,Du:2025xes,DESI:2025ffm,Rodrigues:2025hso,Feng:2025mlo,Feng:2026pzs,Ladeira:2026jne}, the introduction of dark radiation and sterile neutrinos \citep{Allali:2024anb,Benso:2024qrg,Barenboim:2025vrc,Feng:2025mlo,Du:2025iow}, neutrino self-interactions and non-standard interactions \citep{Whitford:2025dmq,Chattopadhyay:2025ccy,Zu:2025lrk}, as well as the utilization of advanced large-scale structure analyses \citep{Ivanov:2024jtl,Racco:2024lbu,Sui:2024wob,Chudaykin:2025lww,Simoes:2025ejj,Chudaykin:2025lww,Labate:2025whw}.

Cosmological measurements of the neutrino mass depend sensitively on the assumed cosmological model and particularly on the dynamical nature of DE \citep{Zhang:2017rbg}. Previous systematic investigations have revealed a strong degeneracy between $\sum m_\nu$ and the EoS of DE parameter $w$ \citep{Zhang:2015uhk,Wang:2016tsz,Zhao:2016ecj,Zhang:2017rbg,Yang:2017amu,Zhao:2017jma,RoyChoudhury:2018gay,Vagnozzi:2018jhn,Zhao:2018fjj,Feng:2019mym,Feng:2019jqa,Zhang:2020mox}. These studies demonstrated that cosmological constraints on the neutrino mass become significantly more stringent than those in $\Lambda$CDM when $w$ evolves from a larger value to a smaller one, such as in quintessence models or specific quintom scenarios. Conversely, scenarios where the EoS evolves from a smaller value to a larger one, including phantom DE models ($w < -1$) or alternative quintom crossings, naturally relax the upper limit on $\sum m_\nu$.

\begin{figure}[htbp]
\centering
\includegraphics[width=0.47\textwidth]{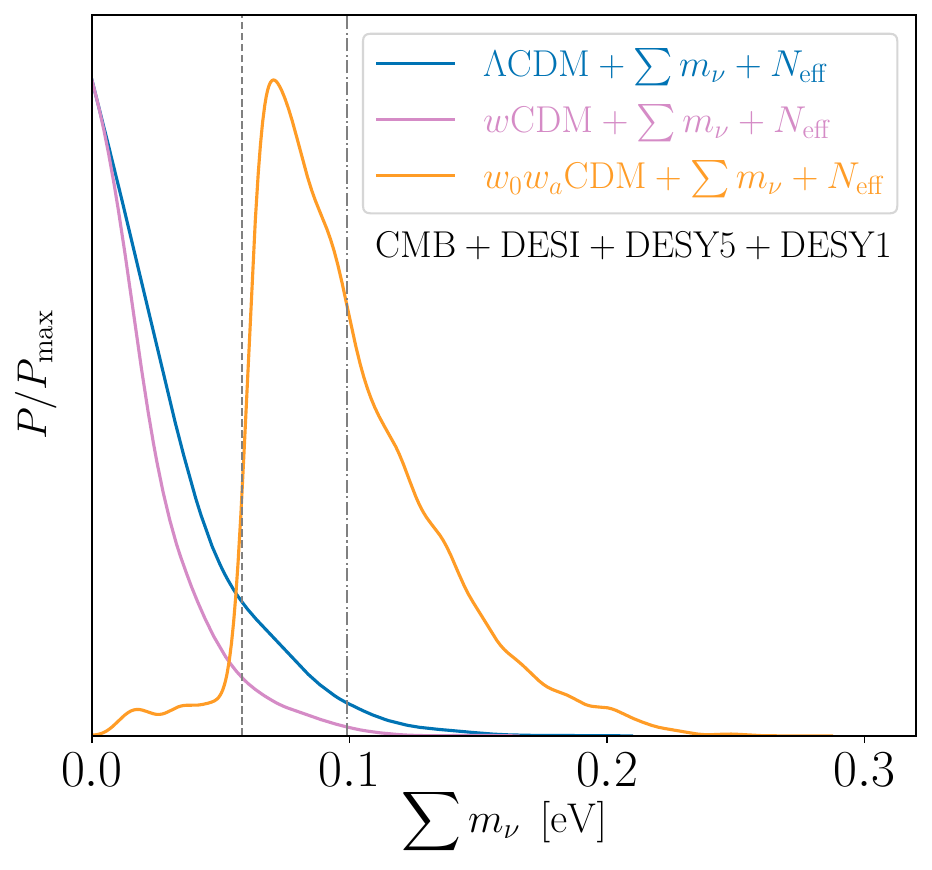}
\caption{The one-dimensional marginalized posterior distributions of $\sum m_{\nu}$ in the $\Lambda\mathrm{CDM}+\sum m_\nu+N_{\rm eff}$, $w\mathrm{CDM}+\sum m_\nu+N_{\rm eff}$, and $w_0w_a\mathrm{CDM}+\sum m_\nu+N_{\rm eff}$ models using the CMB+DESI+DESY5+DESY1 data \citep{Du:2025xes}.}
\label{fig11}
\end{figure}

Therefore, as previously noted, the DDE preferred by current DESI data, characterized by an EoS crossing from $w<-1$ to $w>-1$, can naturally relax the neutrino mass limits and ease this tension. A comprehensive investigation by the DESI collaboration highlighted the capability of DDE to resolve the neutrino mass discrepancy~\citep{Elbers:2025vlz}. To accurately quantify the tension with particle physics experiments, this study introduced a non-physical assumption that allows the effective neutrino mass to be negative (representing cosmological observational effects opposite to those produced by positive-mass neutrinos). Their analysis revealed that under the standard $\Lambda$CDM model, the data prefer a negative neutrino mass, leading to a $3\sigma$ tension with the NH lower limits established by oscillation experiments. Nevertheless, within the CPL framework, this tension is significantly relieved, yielding an upper limit of $\sum m_\nu < 0.163\,{\rm eV}$ and successfully reconciling the cosmological bounds with particle physics experiments.

Building on this, \citet{Du:2025xes} demonstrated that when weak gravitational lensing data are taken into account, simultaneously measuring the neutrino mass and the effective number of relativistic degrees of freedom $N_{\rm eff}$ within the CPL model yields a positive neutrino mass, thereby further alleviating the tension. A joint analysis incorporating DESI DR2 BAO, CMB, DESY5 SN, and DESY1 weak gravitational lensing data yields a total neutrino mass of $\sum m_\nu = 0.098^{+0.016}_{-0.037}\,{\rm eV}$, marking a detection of a positive neutrino mass at the $2.7\sigma$ confidence level that is highly compatible with the NH lower bound, as shown in Figure~\ref{fig11}. Additionally, expanding the parameter space provides complementary avenues to alleviate the tension. For instance, \citet{RoyChoudhury:2025dhe} showed that considering an extended twelve-parameter model, which simultaneously includes the running of the scalar spectral index, the CMB lensing amplitude parameter, and DDE, results in a $\sim 2\sigma$ detection of a positive neutrino mass ($\sum m_\nu = 0.190\pm0.088\,{\rm eV}$).

At the perturbation level, the preference for a negative effective neutrino mass indicates that current matter fluctuations and structure growth may be higher than the expectations of the standard $\Lambda$CDM model. To address this, \citet{Giare:2025ath} introduced a free growth index parameter $\gamma$ to characterize the growth rate of matter perturbations. They found that allowing $\gamma$ to deviate from the standard $\Lambda$CDM expectation ($\gamma \simeq 0.55$) significantly relaxes the neutrino mass bounds. The observational data consistently favor an enhanced structure growth scenario with $\gamma > 0.55$, which completely eliminates the tension without altering the inferred cosmic matter density and elevates the upper bound to $\sum m_\nu \lesssim 0.13 \sim 0.2\,{\rm eV}$.

\begin{figure}[htbp]
\centering
\includegraphics[width=0.48\textwidth]{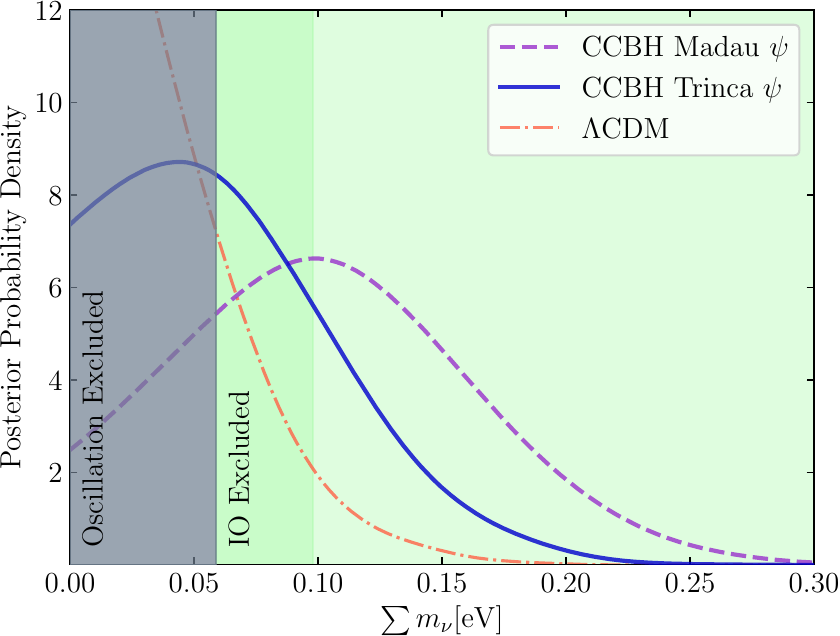}
\caption{The one-dimensional marginalized posterior distributions of $\sum m_{\nu}$ in the CCBH Madau $\psi$ and CCBH Trinca $\psi$ models using the CMB+DESI DR2 BAO data \citep{DESI:2025ffm}.}
\label{fig12}
\end{figure}

An alternative pathway to alleviate the tension involves constructing a physical late-time conversion mechanism where matter is actively transformed into DE, thereby carving out space for a larger neutrino mass without modifying the total energy density. \citet{DESI:2025ffm} analyzed a cosmologically coupled black hole (CCBH) scenario to represent this mechanism. By linking the generation of DE to the cosmic star formation history, the CCBH model consumes baryons to produce DE after the reionization epoch. As shown in Figure~\ref{fig12}, their analysis demonstrates that such matter to energy conversion accurately recovers the cosmological expansion history while yielding a positive neutrino mass of $\sum m_\nu = 0.106^{+0.050}_{-0.069}\,{\rm eV}$ using the CMB+DESI DR2 BAO data, which aligns with neutrino oscillation lower bounds. As an additional consequence, this late-time baryon consumption provides a potential astrophysical explanation for the missing baryons problem.

Finally, the neutrino-mass problem is intimately connected to the optical depth to reionization $\tau$. \citet{Sailer:2025lxj} pointed out that because the CMB inferred $\tau$ is negatively correlated with the matter fraction, assuming a higher optical depth ($\tau \sim 0.09$) can simultaneously resolve the geometric neutrino-mass problem and suppress the preference for DDE. Although such a high value of $\tau$ typically conflicts with large-scale CMB polarization measurements, \citet{Namikawa:2025doa} proposed that cosmic birefringence could reconcile this discrepancy. Specifically, phase ambiguities in the cosmic birefringence angle during reionization can suppress the reionization bump in the E-mode polarization power spectrum, allowing a higher optical depth to be fully consistent with current data.

Overall, the neutrino-mass problem highlighted by recent DESI data demonstrates that cosmological constraints on the absolute neutrino mass are highly sensitive to the assumed theoretical framework. Extending the standard $\Lambda$CDM model through DDE, modified structure growth rates, matter conversion mechanisms, or new reionization physics can effectively resolve this tension. These diverse theoretical approaches provide viable mechanisms to reconcile cosmological observations with particle physics limits and offer complementary perspectives for exploring cosmology beyond the standard paradigm.

\section{Conclusion}\label{sec:conclusion}

The recent release of high-precision DESI BAO measurements has provided a timely opportunity to reassess the late-time expansion history of the universe. When combined with CMB and SN data, these results exhibit a deviation from the $\Lambda$CDM model, favoring DDE models within the CPL framework. This preference typically points toward $w_0 > -1$ and $w_a < 0$, corresponding to a quintom evolutionary scenario wherein DE behaved more like a phantom fluid in the past, while approaching a quintessence state today. If confirmed, this would constitute one of the most compelling indications to date that the simple $\Lambda$CDM model provides an incomplete description of the accelerated cosmic expansion. However, the current observational landscape also warrants caution. The statistical significance of DDE is highly dependent on the adopted $w(a)$ parameterization, the choice of SN datasets, and the calibration of low-redshift distance anchors. Consequently, alongside DDE, it is imperative to explore alternative new physics beyond the $\Lambda$CDM paradigm to comprehensively account for the DESI observations. In this context, the present review focuses on the observational evidence and goodness-of-fit of several new physics frameworks beyond the $\Lambda$CDM model, including interacting DE, non-minimally coupled gravity, and non-standard dark matter.

IDE provides an alternative physical interpretation for DESI observational results. Due to the energy-momentum exchange between DE and dark matter, IDE can naturally mimic DDE at the background level. Several analyses indicate that the observational evidence for IDE can reach up to $5\sigma$ given the current CMB, DESI, and SN data. Furthermore, IDE models can fit current observations just as well as phenomenological DDE parameterization schemes. Investigations into sign-changing interactions and non-parametric reconstructions further illustrate that the deviations revealed by DESI observations may reflect the complex evolution of the dark matter-DE system. This indicates that, beyond the intrinsic dynamical evolution of DE, IDE serves as a viable theoretical alternative to account for the background geometric deviations suggested by DESI.

Non-minimally coupled scenarios provide another important class of explanations. In these models, the apparent evolution of the EoS of DE may be an effective description of additional gravitational degrees of freedom rather than the physical EoS of a separately conserved fluid. For instance, the HTG model successfully reproduces the evolutionary trajectory preferred by DESI data, naturally depicting the physical picture of a smooth transition of the EoS from the phantom regime at intermediate redshifts to the current quintessence regime. This model garners strong Bayesian statistical support ($\ln B = 7.34 \pm 0.6$), demonstrating that non-minimally coupled gravity theories are highly competitive in fitting the latest observations.

The observed deviations of DESI from $\Lambda$CDM could also arise from our conventional assumptions regarding the properties of dark matter. In non-standard dark matter models, which relax the assumption of perfectly pressureless dark matter to allow for a non-zero EoS parameter ($w_{\rm dm}$), current data exhibit a statistical preference of approximately $2.4\sigma$ to $3\sigma$ for a non-zero $w_{\rm dm}$, posing a challenge to the traditional CDM paradigm to some extent. Although this statistical preference depends partly on the specific DE model, it clearly demonstrates that non-standard dark matter serves as an equally viable avenue for interpreting the current DESI data.

These theoretical extensions also play an important role in alleviating the current cosmological tensions. DDE can loosen the upper bound constraints on neutrino masses, effectively mitigating the tension between cosmological observations and the lower limits set by terrestrial particle physics oscillation experiments. Meanwhile, mechanisms such as IDE, modified gravity, and early-universe modifications can theoretically provide potential solutions to alleviate the $H_0$ tension or the $S_8$ tension.

In conclusion, the DESI data compel us to reconsider the foundational assumptions of standard cosmology. While these results may hint at new physics beyond the $\Lambda$CDM model, the precise nature of this new physics remains far from established. The critical question is no longer merely whether the data favor $w(z) \neq -1$, but rather which underlying assumption of the standard cosmological model is being tested: the constancy of DE, the independent conservation of dark matter and DE, the validity of General Relativity on cosmological scales, or the cold and pressureless nature of dark matter. Addressing this issue requires us to move beyond background fitting toward a unified interpretation of geometry, growth, and systematic errors, as well as a unified diagnostic of new physics. To this end, a comprehensive discussion on the unified diagnostic of various new physics scenarios and a multi-probe joint analysis of future background geometry and structural growth will be presented in our review paper currently in preparation (X. Zhang 2026, in preparation).

\section*{Acknowledgments}

We thank Sheng-Han Zhou, Yi-Min Zhang, Hui Li, Zhao-Yu Li, and William Giar\`e for helpful discussions. This work was supported by the National Natural Science Foundation of China (NSFC, grants Nos. 12533001, 12575049, and 12473001), the National SKA Program of China (grants Nos. 2022SKA0110200 and 2022SKA0110203), the China Manned Space Program (grant No. CMS-CSST-2025-A02), and the National 111 Project (grant No. B16009).

\bibliography{DESI}

\end{document}